\documentclass[12pt,english,aps,preprint,floats]{revtex4}
\usepackage[T1]{fontenc}
\usepackage[latin1]{inputenc}
\usepackage{color}
\usepackage{graphicx}
\usepackage{amssymb}

\makeatletter

\usepackage{amssymb,amsmath,graphicx,epsfig}

\newcommand{\bee}{\begin{equation}}
\newcommand{\ee}{\end{equation}}
\newcommand{\beea}{\begin{eqnarray}}
\newcommand{\eea}{\end{eqnarray}}

\usepackage{babel}

\usepackage{babel}
\makeatother
\begin{document}

\title{Potentials for Light Moduli in Supergravity and String Theory}

\author{S. P. de Alwis\protect}

\affiliation{Perimeter Institute, 31 Caroline Street N., Waterloo, ON N2L 2Y5,
Canada}

\affiliation{Department of Physics, University of Colorado, Box 390, Boulder,
CO 80309.\\
 \\
 \texttt{e-mail: dealwis@pizero.colorado.edu} }

\begin{abstract}
This is an account of lectures that were given at TASI 2005,  the
Shanghai Summer School in M-theory 2005 and the Perimeter Institute.
I review 1) the derivation of the potential for chiral scalar fields
in ${\cal N}$=1 supergravity 2) the relation between F and D terms
for chiral scalars, Weyl anomalies and the generation of non-perturbative
terms in the superpotential and 3) the derivation of effective potentials
for light moduli in type IIB string theory.
\end{abstract}
\maketitle

\section{Introduction}

\label{S:intro}

These lectures are aimed at giving a logically consistent account of
recent work on potentials for moduli in string theory. To this end
I have tried to give a systematic presentation of the supergravity
formulae that are at the basis of these discussions and to show how
these potentials will arise in string theory (in particular in type
IIB compactified on Calabi-Yau orientifolds with D-branes and fluxes).
The lectures are organized as follows:

\begin{enumerate}
\item I discuss ${\cal N}=1$ global supersymmetric actions for chiral scalar
superfields in superspace formalism and derive the potential. I motivate
the construction of the corresponding supergravity action coupled
to chiral scalars and derive the F-term potential for the latter. 
\item I discuss the coupling of gauge fields in global and local supersymmetry
and the D-term contribution to the potential. In particular a relation
between the F and D terms is derived. Finally I discuss the issue
of Weyl anomalies and the derivation of the non-perturbative contribution
to the superpotential from gaugino condensation.
\item I discuss the derivation of potentials for moduli in type IIB supergravity
following the work of Giddings, Kachru and Polchinski, and Kachru,
Kallosh, Linde and Trivedi. Then I discuss how heavy moduli may be
integrated out in supergravity theories to give effective theories
for light moduli. 
\end{enumerate}
Sections 1 and 2 depend heavily on two classic text books, Wess and
Bagger \cite{Wess:1992cp} (whose notation and conventions I use also)
and Gates, Grisaru, Rocek and Siegel \cite{Gates:1983nr}. Section
2. also uses the work of \cite{Kaplunovsky:1994fg}\cite{Burgess:1995aa}.
Section 3 begins with a short review of \cite{Giddings:2001yu} and\cite{Kachru:2003aw}.
I should mention here that these works are in turn based on earlier
work on flux compactifications such as \cite{Dine:1985rz}\cite{Strominger:1986uh}\cite{Polchinski:1995sm}\cite{Becker:1996gj}\cite{Dasgupta:1999ss}\cite{Taylor:1999ii}\cite{Gukov:1999ya}\cite{Greene:2000gh}\cite{Grana:2000jj}
even though I do not use these directly. Then I review some of my
own attempts \cite{deAlwis:2003sn}\cite{deAlwis:2004qh}\cite{deAlwis:2005tg}\cite{deAlwis:2005tf}
to understand the logical structure of these derivations. It should
be noted that these lectures represent my personal view on these issues
rather than being a comprehensive review. While they contain references
to other work on these matters the list is far from being complete.
The focus of these lectures is on supergravity and the derivations
of moduli potentials. For comprehensive reviews of the subject of
flux compactifications as such, the reader should consult the recent
reviews \cite{Frey:2003tf}\cite{Balasubramanian:2004wx}\cite{Silverstein:2004id}
\cite{Grana:2005jc}.

\section{Potential for chiral scalars in ${\cal N}=1$ supergravity\label{sec:Potential-for-chiral}}

String phenomenology is based on integrating out string states and
Kaluza-Klein (KK) modes to get an ${\cal N}=1$ supergravity (SUGRA).
The standard argument in the context of superstring theory (which
so far has only an on-shell S-matrix formulation)) is that one needs
to compactify the extra six dimensions of string theory, on a manifold
which admits one killing spinor. This ensures that the four dimensional
theory has supersymmetric vacua. The fluctuations around such a vacuum
must necessarily be described by four dimensional supergravity, and
the requirement that there is one and only one killing spinor, guarantees
${\cal N}=1$ four dimensional supersymmetry. Once one has derived
this $ $ SUGRA it is possible to argue that the non-supersymmetric
solutions (if they exist) are also (low energy) solutions to string
theory. This is the approach that is normally taken in string phenomenology
and that is what we will adopt. 

We will only discuss ${\cal N}=1$ supersymmetry in these notes so
from now on this should understood. The notation and conventions are
as in Wess and Bagger \cite{Wess:1992cp}.

\subsection{Global supersymmetry in Superspace}

Global supersymmetry is defined by a Weyl spinor supercharge $Q_{\alpha}\,\alpha=1,2$
and its complex conjugate $\bar{Q}{}_{\dot{\alpha}}$ that transform
bosons into fermions. The most convenient formulation of supersymmetry
is in terms of superfields which may be viewed as fields that live
in a space (superspace) with superspace coordinate $z^{M}=\{ x^{m},\theta^{\mu},\theta_{\dot{\mu}}\}$with
$x^{m},\, m=0,..3$ being the usual space-time coordinates and $\theta^{\mu},\,\mu=1,2$
(and its complex conjugate) being fermionic coordinates represented
by Grassman numbers. The supercharge is then essentially a translation
operator in the fermionic direction $Q_{\alpha}=\partial_{\alpha}-i\sigma_{\alpha\dot{\alpha}}^{m}\bar{\theta}^{\dot{\alpha}}\partial_{m}$
(with $\partial_{\alpha}=\partial/\partial\theta_{\alpha},\:\partial_{m}=\partial/\partial x^{m}$and
$\sigma^{m}$are the Pauli matrices for $m=1,2,3,$ and the unit matrix
for $m=0.$ The conjugate supercharge is $\bar{Q}_{\dot{\alpha}}=-\partial_{\dot{\alpha}}+i\theta^{\alpha}\sigma_{\alpha\dot{\alpha}}^{m}\partial_{m}$.
The supercharges satisfy the algebra\begin{eqnarray}
\{ Q_{\alpha},\bar{Q}{}_{\dot{\alpha}}\} & = & 2i\sigma_{\alpha\dot{\alpha}}^{m}\partial_{m}\label{susyalg1}\\
\{ Q_{\alpha},Q_{\beta}\}=0 &  & \{\bar{Q}{}_{\dot{\alpha}},\bar{Q}{}_{\dot{\beta}}\}=0\label{susyalg2}\end{eqnarray}
A general superfield has many terms and is a reducible representation
of supersymmetry. The simplest irreducible representation is the chiral
scalar superfield which in a sense depends only on the thetas and
not on their complex conjugates. To make this statement in a supersymmetric
way we introduce the operator $D_{\alpha}=\partial_{\alpha}+i\sigma_{\alpha\dot{\alpha}}^{m}\bar{\theta}^{\dot{\alpha}}\partial_{\dot{\alpha}}$
such that it and its conjugate anti-commute with all the super charges
(and give another representation of the superalgebra). Thus a chiral
scalar superfield $\Phi$ is defined by the constraint $ $\begin{equation}
\bar{D}_{\dot{\alpha}}\Phi=0,\label{chiral}\end{equation}

whose solution is (in terms of a bosonic field $A(x)$ a fermionic
field $\psi_{\alpha}(x)$ and an auxiliary (i.e. non-propagating)
field $F(x)$ is,

\[
\Phi=A(y)+\sqrt{2}\theta^{\alpha}\psi_{\alpha}(y)+\theta^{\alpha}\theta_{\alpha}F(y)\]
where $y=x^{m}+i\theta\sigma^{m}\bar{\theta}$. The conjugate of this
chiral superfield would be anti-chiral. Note that fermionic indices
are raised and lowered with $\epsilon_{\alpha\beta}:\,\epsilon_{21}=\epsilon^{12}=1$.
Supersymmetry transformations take $A$ to $\psi$ the latter to $F$.
The highest component of a superfield (in this case $F$) transforms
into a total derivative. Also note that since $D^{3}=\bar{D}^{3}=0$,
the operator ($\bar{D}^{2}$)$\bar{D}^{2}$ is a (anti-) chiral projector.

The next simplest irreducible representation is the real superfield
($V$) defined by $ $$ $\begin{equation}
V=V^{\dagger},\:{\rm giving\, components}\{ c,\chi^{\alpha},\bar{\chi}^{\dot{\alpha}},M,N,\lambda^{\alpha},\lambda^{\dot{\alpha}},v_{m},D\}.\label{realsupfield}\end{equation}

Note that the definition is ambiguous up to Kaehler gauge transformations\begin{equation}
V\rightarrow V+\Lambda+\Lambda^{\dagger}\label{kaehlergauge}\end{equation}
where $\Lambda$ is a chiral superfield. Looking at the transformations
of the components it is easy to see that the first five components
are gauge degrees of freedom. In a particular gauge which still leaves
the freedom to make the usual gauge transformations (the so-called
Wess-Zumino (WZ) gauge) a real super field has a fermion (so-called
gaugino) its conjugate, a real vector field (these are physical propagating
fields) and a scalar auxiliary field $D$. Again the highest component,
i.e. $D$, transforms into a total derivative. It is often convenient
to redefine components in terms of the $D_{\alpha}$ operator. Thus
for a chiral superfield the scalar the fermion and the auxiliary $F$
field may be defined by,

\begin{equation}
A=\Phi|,\,\psi_{\alpha}=\frac{1}{\sqrt{2}}D_{\alpha}\Phi|,\, F=-\frac{1}{4}D^{\alpha}D_{\alpha}\Phi|\label{components}\end{equation}

where the vertical bar is an instruction to set $\theta=\bar{\theta}=0$
after performing the operations on the superfield. Also the D-term
in a real field would now be defined by $D=(-\frac{1}{4}\bar{D}_{\dot{\alpha}}\bar{D}^{\dot{\alpha}})(-\frac{1}{4}D_{\alpha}D^{\alpha})V|$.
The product of two chiral superfields would also be chiral while the
product of a chiral field and its conjugate would be a real superfield.
Note also that under the space-time integral sign,

\begin{equation}
\int d^{2}\theta\rightarrow-\frac{1}{4}D_{\alpha}D^{\alpha},\,{\rm and}\int d^{4}\theta\equiv\int d^{2}\theta d^{2}\bar{\theta}\rightarrow\frac{1}{16}\bar{D}^{2}D^{2}\label{replace}\end{equation}

We may now write the most general globally supersymmetric action for
chiral superfields as\begin{equation}
S=\int d^{4}x\left[\int d^{4}\theta K(\Phi,\bar{\Phi})+\left(\int d^{2}\theta W(\Phi)+h.c.\right)\right]\label{susyaction}\end{equation}

In the above $W$ the so-called superpotential is a holomorphic function
of $\Phi$ and so is a chiral field whilst $K$ the so-called Kaehler
potential is a real superfield. For example $W$ could be a polynomial
in $\Phi$ while $K=\Phi\bar{\Phi}.$ The integrals over the thetas
are essentially instructions to pick the D or the F terms of the corresponding
integrals and since these transform into total derivatives the action
above is invariant under supertransformations. Clearly in the above
we may replace the superfield $\Phi$ by a set of superfields $\{\Phi^{i}\}$.

The action also has invariance under Kaehler transformations, \begin{equation}
K(\Phi,\bar{\Phi})\rightarrow K(\Phi,\bar{\Phi)}+\Lambda(\Phi)+\bar{\Lambda}(\bar{\Phi}),\label{kaehler1}\end{equation}

where $\Lambda$ is an arbitrary chiral superfield.

Let us compute the potential for scalar fields from (\ref{susyaction}).
Retaining just the scalar contributions and using (\ref{replace})(\ref{components})
we have\begin{eqnarray}
\int d^{4}\theta K\rightarrow K_{i\bar{j}}F^{i}F^{\bar{j}}, & \int d^{2}\theta W\rightarrow F^{i}W_{i} & \int d^{2}\bar{\theta}\bar{W}\rightarrow\bar{F}^{\bar{i}}\bar{W}_{\bar{i}}.\label{potreplace}\end{eqnarray}

Hence we have\begin{equation}
{\cal -V}=K_{i\bar{j}}F^{i}F^{\bar{j}}+F^{i}W_{i}+\bar{F}^{\bar{i}}\bar{W}_{\bar{i}},\label{susypot1}\end{equation}

where $W_{i}=\partial W/\partial\Phi^{i}$ and $K_{i\bar{j}}=\partial_{i}\partial_{\bar{j}}K$. 

The auxiliary field has an algebraic equation of motion - $\partial{\cal V}/\partial F^{i}=K_{i\bar{j}}F^{\bar{j}}+W_{i}=0$.
Hence we may rewrite the potential as\begin{equation}
{\cal V}=K_{i\bar{j}}F^{i}\bar{F}^{\bar{j}}=K^{i\bar{j}}W_{i}\bar{W}_{\bar{j}}\label{susypot2}\end{equation}

It is useful to note that the potential can be computed either from
evaluating just the F-term in the action (i.e. the $\int d^{2}\theta$
term) or its conjugate, or the D-term (the $\int d^{4}\theta$ term)
with reversed sign.

Let us now compute the superfield equation of motion. To do this it
is convenient to use the first replacement in (\ref{replace}) to
write the action as \begin{equation}
S=\int d^{4}x\int d^{2}\theta[-\frac{1}{4}\bar{D}^{2}K(\Phi,\bar{\Phi})+W(\Phi)]+\int d^{2}\bar{\theta}\bar{W}(\bar{\Phi}).\label{susyaction2}\end{equation}

When this action is varied w.r.t. the chiral field, the variation
can be pulled through the operator $\bar{D}$, so that we immediately
get the field equation,$ $\begin{equation}
\frac{1}{4}\bar{D}^{2}K_{i}=W_{i}\label{eom1}\end{equation}
which may be rewritten as \begin{equation}
\frac{1}{4}K_{i\bar{j}}\bar{D}^{2}\bar{\Phi}^{\bar{j}}+\frac{1}{4}K_{i\bar{j}\bar{l}}\bar{D}_{\dot{\alpha}}\bar{\Phi}^{\bar{j}}\bar{D}^{\dot{\alpha}}\bar{\Phi}^{\bar{l}}=\frac{\partial W}{\partial\Phi^{i}}\label{eom2}\end{equation}

The lowest component of this equation gives the auxiliarly field equation
$\bar{F}^{\bar{j}}=-K^{\bar{j}i}W_{i}+fermion\, terms$. Operating
on it once with $\bar{\textrm{D}}_{\dot{\alpha}}$ and then setting
the thetas to zero gives the fermion equation and doing this twice
gives the bosonic equation\begin{equation}
\partial^{2}\bar{\Phi}^{\bar{j}}=K^{\bar{j}i}\frac{D^{2}}{4}W_{i}+\ldots=-K^{\bar{j}i}W_{il}F^{l}+\ldots=K^{\bar{j}i}\frac{\partial{\cal V}}{\partial\Phi^{i}}+\ldots,\label{bosoneom}\end{equation}

where the ellipses represent terms which are quadratic in the fermions
and we've used the identity\[
\frac{1}{16}D^{2}\bar{D}^{2}\bar{\Phi}=\partial^{2}\bar{\Phi}.\]

\subsection{Supergravity in Superspace}

The global SUSY action is invariant under the constant SUSY transformations
$ $\begin{equation}
z^{M}\rightarrow z^{M}-\xi^{M},\, z^{M}\equiv\{ x^{m},\theta^{\mu},\theta_{\dot{\mu}}\}\label{globaltrans}\end{equation}

Local SUSY transformations naturally lead to general coordinate transformations
since the anti-commutator of two supercharges generate translations,
which in this case become general coordinate transformations. Thus
the above transformations are replaced by the following general coordinate
transformations (GCT) in superspace:\begin{eqnarray}
x^{m} & \rightarrow & -i(\theta\sigma^{m}\bar{\xi}(x)-\xi(x)\sigma^{m}\bar{\theta})\label{gct}\\
\theta^{\mu}\rightarrow\theta^{\mu}-\xi^{\mu}(x), &  & \bar{\theta}_{\dot{\mu}}\rightarrow\bar{\theta}_{\dot{\mu}}-\bar{\xi}_{\dot{\mu}}(x)\label{localsusy}\end{eqnarray}
or $z^{M}\rightarrow z'^{M}-\xi^{M}(z)$.

We need to to introduce a set of frames with Lorentz superspace indices
$A=\{ a,\alpha,\dot{\alpha}\}$%
\footnote{These frame indices are often called flat indices in the supergravity
literature whilst the index set $M$ are called curved indices)%
} with the supervielbein one-form in superspace being written as\begin{equation}
E^{A}=dz^{M}E_{M}^{A}(z);\, E_{M}^{A}E_{A}^{N}=\delta_{M}^{N};\, E_{A}^{M}E_{M}^{B}=\delta_{A}^{B}.\label{vielbein}\end{equation}

These are taken to have the following transformation properties:

\begin{equation}
\delta_{\xi}E_{M}^{A}=-\xi^{N}\partial_{N}E-(\partial_{M}\xi^{N})E_{N}^{A}\label{Egct}\end{equation}

under superspace GCT, and \[
\delta E_{M}^{A}=E_{M}^{B}L_{B}^{\,\, A}(z),\,{\rm where}L_{B}^{\,\, A}=\{ L_{b}^{\,\, a},L_{\beta}^{\,\,\alpha},L_{\,\dot{\,\alpha}}^{\dot{\beta}}\}\]

(with the different L's being matrices in the Lorentz algebra in the
vector and Weyl (anti) spinor representations. One may also define
a Lorentz connection one form\begin{equation}
{\bf \phi}=dz^{M}{\bf \phi}_{M}\,{\rm with\,}{\bf \phi}_{M}=\{\phi_{MB}^{\,\,\,\,\,\,\,\,\, A}\}\label{connection}\end{equation}

which transforms as \begin{equation}
\delta\phi=\phi L-L\phi-dL\label{Lorconnection}\end{equation}

under Lorentz transformations. Also one defines a torsion superfield
two-form by,$ $\begin{equation}
T^{A}=\nabla E^{A}=dE^{A}+E^{B}\phi_{B}^{\,\, A}=\frac{1}{2}dz^{M}dz^{N}T_{NM}^{\,\,\,\, A}=\frac{1}{2}E^{C}E^{B}T_{BC}^{\,\,\,\, A}.\label{torsion}\end{equation}

It is important to realize that torsion on superspace does not vanish
even if the metric is flat. Thus in flat space we have the super-covariant
derivative with $D_{A}=e_{A}^{\,\, N}\partial_{N}$. The inverse of
the matrix defines the flat space vielbein one form by $E^{A}=e_{M}^{A}dz^{M}$with
$ $$e_{a}^{m}=\delta_{a}^{m},\, e_{\alpha}^{m}=i\sigma_{\alpha\dot{\alpha}}^{m}\bar{\theta}^{\dot{\alpha}}\ldots$.
In flat superspace there exist transformations to coordinates in which
$\phi=0$ but torsion has non-vanishing components \begin{equation}
T_{\alpha\dot{\beta}}^{c}=T_{\dot{\beta}\alpha}^{c}=2i\sigma_{\alpha\dot{\beta}}^{c}.\label{flattorsion}\end{equation}

As in ordinary differential geometry here too one can define a curvature
two form by\begin{equation}
{\cal R}_{B}^{\,\, A}=d\phi_{B}^{\,\, A}+\phi_{B}^{\,\, C}\wedge\phi_{C}^{\,\, A}\label{curvature}\end{equation}

Thus one needs to solve\begin{equation}
\nabla\nabla E^{A}=\nabla T^{A}=E^{B}{\cal R}_{B}^{\,\, A}\label{curvature2}\end{equation}

subject to torsion constraints such as (\ref{flattorsion}) except
that (unlike in flat space) not all of the other components vanish.
The result is that the minimal number of independent components of
the supervielbein are\begin{equation}
e_{m}^{a},\,\psi_{m}^{\alpha},\,\bar{\psi}_{m\dot{\alpha}}\,,M,\, b_{a}=\bar{b}_{a}\label{chiralR}\end{equation}
where the first is the usual vielbein, the second and third are the
gravitino and its conjugate, and the last two are a complex scalar
auxiliary field and a real vector auxiliarly field. In addition it
turns out that all components of the torsion and the curvature superfields
can be expressed in terms of three superfields\begin{equation}
R,\,\, G_{\alpha\dot{\alpha}},\,\, W_{\alpha\beta\gamma}\label{sugrasuperfields}\end{equation}

(with the last being symmetric in its indices) subject to constraints
such as the covariant chirality constraint\begin{equation}
\bar{\nabla}_{\dot{\alpha}}R=0.\label{chiralR2}\end{equation}

In SUGRA the chirality projector $\bar{D}^{2}$ of global supersymmetry
is replaced by\begin{equation}
\bar{\nabla}_{\dot{\alpha}}\bar{\nabla}^{\dot{\alpha}}-8R.\label{chiralproj}\end{equation}

It turns out also that$ $\begin{equation}
R(z)|=-\frac{1}{6}M(x)\ldots,\nabla^{\alpha}\nabla_{\alpha}R(z)|=\frac{1}{3}e_{a}^{m}e_{b}^{n}{\cal R}_{mn}^{\,\,\,\, ab}(x)+\ldots,,G_{a}(z)|=-\frac{1}{3}b_{a}(x).\label{lowestcomps}\end{equation}

\subsection{General Matter SUGRA Action}

Now we can discuss the action for supergravity coupled to chiral scalar
matter. It is useful to rewrite the supervielbein $E_{m}^{a}$in
terms of a real axial vector field superfield $H_{m}$, and a chiral
superfield $\varphi$ (called the chiral compensator) which satisfies
the $flat\, space$ chirality constraint $\bar{D}_{\dot{\alpha}}\varphi=0.$
It turns out that the graviton and the gravitino as well as the axial
vector auxiliary field (in Wess-Zumino gauge) are contained in $H_{m}$.
Also the superdeterminant%
\footnote{For a definition and more details see  section 3.7  and chapter 5 of \cite{Gates:1983nr}.
Here we will only need to know that they are superdensities which
enable one to define superspace integrals which are invariant under
local SUSY transformations. %
} of the supervielbein may be written as\begin{equation}
{\bf E}\equiv{\rm sdet}E_{M}^{A}={\cal E}(H)\bar{\varphi}\varphi\label{superdet}\end{equation}

For any superfield ${\bf L}$ there is a useful identity that relates
different expressions for superfield actions found in the literature,\begin{equation}
-\frac{1}{4}\bar{D}^{2}{\bf EL}=\varphi^{3}(-\frac{1}{4}\bar{\nabla}^{2}+2R){\bf L.}\label{identity}\end{equation}

Setting ${\bf L}=R^{-1}$we get since $\bar{\nabla}R=0,$ \begin{equation}
\varphi^{3}=-\frac{1}{4}\bar{D}^{2}\frac{E}{2R}\label{phicubed}\end{equation}

consistent with the fact that the RHS is flatspace chiral. Also setting
${\bf L}=W(\Phi)/2R$, where $W$is a holomorphic function of the
covariantly chiral scalar superfield $\Phi:\,\bar{\nabla}_{\dot{\alpha}}\Phi=0$,
we get\begin{equation}
\varphi^{3}W=-\frac{1}{4}\bar{D}^{2}\frac{{\bf E}}{2R}W(\Phi).\label{identity2}\end{equation}

This identity gives us two equivalent ways of writing the SUGRA invariant
term corresponding to the superpotential term in (\ref{susyaction})\begin{equation}
\int d^{4}xd^{4}\theta\frac{{\bf E}}{2R}W(\Phi)=\int d^{4}xd^{2}\theta\varphi^{3}W(\Phi).\label{Waction}\end{equation}

The chiral density $\varphi^{3}$ is a measure for integrating (over
half of superspace) any chiral term in the action.

The general SUGRA - chiral scalar action can then be written as\begin{eqnarray}
S & = & -\frac{3}{\kappa^{2}}\int d^{8}z{\bf E}(H,\varphi)e^{-\frac{\kappa^{2}}{3}K(\Phi,\bar{\Phi})}\nonumber \\
 &  & +\left(\int d^{6}z\varphi^{3}W(\Phi)+h.c.\right)\label{sugraaction}\end{eqnarray}

In the above we've defined the superspace measures $d^{8}z\equiv d^{4}xd^{4}\theta$
and $d^{6}z\equiv d^{4}xd^{2}\theta$ and $\kappa^{2}\equiv8\pi G_{N}$.
Let us motivate this: First observe that in the flat space limit $\kappa\rightarrow0,\,\varphi\rightarrow1,\,{\bf E\rightarrow1}$,
this reduces to the globally supersymmetric action (\ref{susyaction}).
As in that case $W(\Phi)$, the superpotential, is a holomorphic function
of the chiral superfields while $K(\Phi,\bar{\Phi})$ is a real superfield
which is a function of the chiral fields and the anti-chiral fields.
Secondly let us observe that the superfieldbein determinant contains
the Einstein action.\begin{equation}
-\frac{3}{\kappa^{2}}\int d^{8}z{\bf E}=-\frac{3}{\kappa^{2}}\int d^{6}z\varphi^{3}2R=\frac{2}{\kappa^{2}}\int d^{4}xe{\cal R}+\ldots\label{Einaction}\end{equation}

In the first equality above we used the identity (\ref{identity})
with ${\bf L}=1$, and in the second we've used $\phi^{3}|=e$ and
$-\frac{1}{4}\nabla^{2}R|=-\frac{1}{12}{\cal R}$ where $e=\det[e_{m}^{a}]$
and ${\cal R}$ is the Ricci-scalar (see (\ref{lowestcomps})).

\subsection{Weyl Invariant Formalism\label{ssweyl}}

Before we go on to discuss the potential is useful to consider a slightly
generalized formalism where the Weyl invariance becomes manifest.
We introduce a chiral scalar%
\footnote{This should not to be confused with the connection! In fact we will
not use the latter explicitly in the rest of these notes.%
} $\phi$ (with $\bar{\nabla}_{\dot{\alpha}}\phi=0$). The action (\ref{sugraaction})
(we will set $\kappa^{2}=1$ from now on) is  generalized to\begin{eqnarray}
S & = & -3\int d^{8}z{\bf E}\phi\bar{\phi}e^{-K/3}+\left(\int d^{8}z\frac{{\bf E}}{2R}\phi^{3}W(\Phi)+h.c.\right)\nonumber \\
 & = & -3\int d^{6}z\varphi^{3}(-\frac{1}{4}\bar{\nabla}^{2}+2R)\phi\bar{\phi}e^{-K/3}+\left(\int d^{6}z\varphi^{3}\phi^{3}W(\Phi)+h.c.\right)\label{weylaction}\end{eqnarray}

This action is invariant under Weyl transformations (with a Weyl transformation
parameter chiral superfield $\tau:\,\bar{\nabla}_{\dot{\alpha}}\tau=0$)
given below.\begin{eqnarray}
\Phi\rightarrow\Phi,\,\phi\rightarrow e^{-2\tau}\phi, & \, & \varphi\rightarrow e^{2\tau}\varphi,\, E_{M}^{\alpha}\rightarrow e^{(2\bar{\tau}-\tau)}(E_{M}^{\alpha}-\ldots)\nonumber \\
E_{M}^{a}\rightarrow e^{(\tau-\bar{\tau})}E_{M}^{a},\,{\bf E}\rightarrow e^{2(\tau+\bar{\tau})}{\bf \tau} & \, & \nabla_{\alpha}\rightarrow e^{(\tau-2\bar{\tau})}(\nabla_{\alpha}-\ldots)\label{weyl}\end{eqnarray}

The omitted terms in the above are proportional to the covariant spinor
derivatives of $\tau$.

These transformations are actually an invariance of the torsion and
chirality constrains of minimal supergravity. The Weyl compensator
ensures that the action is also invariant.

The action is also invariant under Kaehler transformations\begin{equation}
K\rightarrow K+f((\Phi,\phi)+\bar{f}(\bar{\Phi},\phi),\, W\rightarrow e^{-f(\Phi,\phi)}W,\,\phi\rightarrow e^{f(\Phi,\phi)/3}\phi\label{kaehler2}\end{equation}

Note that unlike in the case of global SUSY here the superpotential
is not invariant under Kaehler transformations. However the quantity\begin{equation}
G=K+\ln|W|^{2}\label{G}\end{equation}
 is invariant and in fact Kaehler invariance implies that the action
is only dependent on this Kaehler invariant combination and not separately
on $K$ and $W$. It also means that only the zeroes and singularities
of $W$ have an invariant significance. Away from them it can be transformed
to unity.

\subsection{Calculating the Potential}

To discuss the potential it is sufficient to look at conformally flat
metrics $g_{mn}=\sigma^{2}\eta_{mn}$. This amounts to ignoring the
complications coming from the fields $H_{m}$ and therefore in effect
also replacing $\nabla_{\alpha}\rightarrow D_{\alpha}$. In other
words for the purpose of deriving the potential%
\footnote{For details see chapter 8 of \cite{Gates:1983nr}.%
} one may just consider the action,\begin{equation}
S=-\frac{3}{\kappa^{2}}\int d^{8}z\bar{\phi}\phi e^{-\frac{\kappa}{3}K(\Phi,\bar{\Phi})}+\left(\int d^{6}z\phi^{3}W(\Phi)+h.c.\right)\label{actionsimpl}\end{equation}

Note that in the above we have chosen to fix the Weyl gauge by putting
$\varphi=1$. Also $\Phi$ represents a set of chiral scalr superfields
$\{\Phi^{i}\}$. Alternatively we could have chosen $\phi=1$ in which
case the above would have been rewritten in terms of $\varphi$. The
Weyl invariance implies that we can switch the one for the other.
We can derive the equations of motion from this action following the
same procedure as in the global SUSY case (see discussion around (\ref{eom1},
\ref{eom2})). We get (after setting $\kappa=1$)\begin{eqnarray}
-\frac{1}{4}\bar{D}^{2}(\bar{\phi}e^{-K/3}) & = & \phi^{2}W\label{phieom}\\
-\frac{1}{4}\bar{D}^{2}(\phi\bar{\phi}e^{-K/3}K_{i}) & = & -\phi^{3}W_{i}.\label{Phieom}\end{eqnarray}

In the above the subscript $i$ denotes differentiation with respect
to the ith chiral scalar field $\Phi^{i}$. Ignoring fermionic terms
($D_{\alpha}\Phi,\, D_{\alpha}\phi$ etc) we get by taking the lowest
components of the above equations (with $F_{\phi}=-(1/4)D^{2}\phi$
etc.)\begin{eqnarray}
\bar{F}_{\bar{\phi}}e^{-K/3}|-\bar{\phi}\frac{1}{3}K_{\bar{i}}F^{\bar{i}}e^{-K/3}| & = & \phi^{2}W|\label{eq:}\\
\phi[-\frac{1}{4}\bar{D}^{2}(\bar{\phi}e^{-K/3})]K_{i}|+\phi\bar{\phi}e^{-K/3}K_{i\bar{j}}\bar{F}^{\bar{j}} & = & -\phi^{3}W_{i}.\end{eqnarray}

Solving for the two F-terms we have,\begin{eqnarray}
\bar{F}_{\bar{\phi}} & = & e^{K/3}\phi^{2}W+\frac{1}{3}\bar{\phi}K_{\bar{i}}\bar{F}^{\bar{i}}\label{Fphi}\\
\bar{F}^{\bar{j}} & = & -\frac{\phi^{3}}{\phi\bar{\phi}}e^{K/3}K^{i\bar{j}}D_{i}W\label{FPhi}\end{eqnarray}

where we've defined the Kaehler covariant derivative$ $\begin{equation}
D_{i}W=W_{i}+K_{i}W.\label{kaehlerder}\end{equation}

The potential may now be calculated from the superpotential term in
the action (\ref{actionsimpl}) - see discussion around (\ref{susypot2})\begin{eqnarray}
-{\cal V} & = & \int d^{2}\theta\phi^{3}W|_{boson}=-\frac{1}{4}D^{2}(\phi^{3}W)|_{boson}\label{p1}\\
 & = & 3\phi^{2}F_{\phi}W+\phi^{3}W_{i}F^{i}.\label{p2}\end{eqnarray}

Using the expressions (\ref{Fphi},\ref{FPhi}) we get\begin{equation}
{\cal V}=\phi^{2}\bar{\phi}^{2}e^{K/3}(K^{i\bar{j}}D_{i}WD_{\bar{j}}\bar{W}-3|W|^{2})\label{p3}\end{equation}

This potential is not quite in the canonical form since we are not
in the Einstein frame. Recall that the gravitational action is hidden
in the term \begin{equation}
-3\int d^{8}z\phi\bar{\phi}{\cal E}(H)e^{-K/3}\label{einterm}\end{equation}

To get the canonical form of the Einstein action (see \ref{Einaction})
we need to choose \[
\phi|=\bar{\phi|}=e^{K/6}|\]

Then we finally get\begin{equation}
{\cal V}=e^{K}(K^{i\bar{j}}D_{i}WD_{\bar{j}}\bar{W}-3|W|^{2})\label{Fpot1}\end{equation}

This is the well-known F-term potential for chiral scalars coupled
to supergravity first derived in \cite{Cremmer:1978hn}.

We may rewrite this in terms of the Kaehler invariant potential $G$
as (\ref{G})\begin{equation}
{\cal V}=e^{G}(G^{i\bar{j}}G_{i}G_{\bar{j}}-3).\label{Fpot2}\end{equation}

\subsection{Gauge Fields\label{ssec:guage}}

Usually the Kaehler potential (metric) has isometries and if they
are gauged we need to include gauge fields in the theory. A space-time
dependent isometry of the Killing potential (for the moment consider
just linear transformations) is represented on the chiral scalars
by\begin{equation}
\Phi\rightarrow e^{i\Lambda}\Phi,\,\bar{\Phi}\rightarrow\bar{\Phi}e^{-i\bar{\Lambda}},\label{gaugePhi}\end{equation}

(where $\Lambda$ is a chiral scalar superfield) is gauged by introducing
a real superfield $V,\, V^{\dagger}=V,$ which transforms as follows.\begin{equation}
e^{V}\rightarrow e^{i\bar{\Lambda}}e^{V}e^{-i\Lambda}\label{gauge}\end{equation}

Note that in the above we've taken $\Phi$ to be a column matrix in
some representation and $V$ a square matrix $V=V^{a}T_{a}$ where
$T_{a}$ are the generators of the gauge group. Note that we need
two copies of the gauge group ${\cal G}\rightarrow{\cal G}\times{\cal G}$,
since the gauge transformations are represented by a chiral superfield.
As in the Abelian case one can go to the Wess-Zumino gauge in which
\begin{equation}
V=\{ A_{m},\lambda_{\alpha},D\},V^{3}=0.\label{gauge2}\end{equation}

Here $A_{m}$is the usual gauge field, $\lambda$ is its fermionic
partner the gaugino and $D$ is an auxiliary field. Observe that under
gauge transformations\begin{equation}
\bar{\Phi}e^{V}\rightarrow\bar{\Phi}e^{V}e^{-i\Lambda},\label{Phibartrans}\end{equation}
so that invariants are constructed out of $\Phi$ and $\bar{\Phi}e^{V}$.
Thus for instance the globally gauge invariant Kaehler potential $\bar{\Phi}\Phi$
is to be replaced by $\bar{\Phi}e^{V}\Phi$. In general we replace\begin{equation}
K(\Phi,\bar{\Phi})\rightarrow K(\Phi,\bar{\Phi}e^{V}).\label{gaugeinvK}\end{equation}
The gauge field strength is given by a chiral superfield\begin{equation}
{\cal W}_{\alpha}=-(\frac{1}{4}\bar{\nabla}^{2}-2R)e^{-V}\nabla_{\alpha}e^{V}={\cal W}_{\alpha}^{a}T_{a}\label{fieldstrength}\end{equation}
(note that the first factor is the chiral projector) whose components
in the WZ gauge are ${\cal W}_{\alpha}=\{\lambda_{\alpha},F_{mn},D\}$.
The kinetic terms for the gauge fields may then be written as\begin{equation}
\int d^{4}xd^{4}\theta\frac{{\bf E}}{2R}(\frac{1}{4}f_{ab}{\cal W}^{a\alpha}{\cal W}_{\alpha}^{b}+h.c.),\label{gaugeaction}\end{equation}
where in general the gauge coupling function $f_{ab}$, an invariant
tensor of the gauge group, may be a function of the chiral fields
which are neutral under the group. The bosonic part of this action
contains the usual gauge field kinetic term as well as the axion coupling
term, \[
-\frac{1}{4}\int d^{4}x\sqrt{g}(\Re f_{ab}F_{mn}^{a}F^{bmn}-\frac{1}{2}\Im f_{ab}\epsilon^{mnpq}F_{mn}^{a}F_{pq}^{b})\]
Note that with the Weyl transformation rule $V\rightarrow V$ which
implies ${\cal W}\rightarrow e^{-3\tau}W$, the action (\ref{gaugeaction})
is Weyl invariant without any Weyl compensator factor.

To derive the equations of motion for the gauge field from (\ref{gaugeaction})
we use the following trick (for simplicity we'll just consider the
global SUSY case). Define $\Delta V\equiv e^{-V}\delta e^{V}=\delta V+...=\Delta V^{a}T^{a}$
and the gauge covariant derivative $ $$\nabla^{\alpha}\equiv e^{-V}D^{\alpha}e^{V}$.
Then \begin{eqnarray*}
\delta_{V}\frac{1}{2}\int d^{6}zf_{ab}{\cal W}^{\alpha a}{\cal W}_{\alpha}^{b} & = & \int d^{6}zf_{ab}\delta_{V}tr(T^{a}e^{-V}D^{\alpha}e^{V}){\cal W}_{\alpha}^{b}\\
 & = & \int d^{6}zf_{ab}tr(T^{a}[e^{-V}D^{\alpha}e^{V},\Delta V]{\cal W}_{\alpha}^{b}\\
 & = & \int d^{6}zf_{ab}(\nabla^{\alpha}\Delta V)^{a}{\cal W}_{\alpha}^{b}=-\int d^{6}z\Delta V^{a}\nabla^{\alpha}(f_{ab}{\cal W_{\alpha}}^{b}),\end{eqnarray*}
 giving the contribution $ $$\nabla^{\alpha}(f_{ab}{\cal W_{\alpha}}^{b})$
to the gauge field EOM. Also note that under an infinitesimal gauge
transformation, $\Delta_{\Lambda}V=-i\Lambda+ie^{-V}\bar{\Lambda}e^{V}\equiv-i\Lambda+i\tilde{\Lambda}.$

Now we are in a position to evaluate the D-term contribution to the
potential for chiral scalars. As before in evaluating the potential
contributions we ignore curvature terms and Lorentz connection terms.
We make the Kaehler potential term gauge invariant by writing $K(\Phi,\bar{\Phi})\rightarrow K(\Phi,\bar{\Phi}e^{V})+\xi trV)$,
where we've also added a Fayet-Illiopoulos (FI) term when the gauge
group contains a U((1) generator. So the equations of motion for the
auxiliary fields and hence the potential can be derived from\begin{equation}
S=-3\int d^{8}z\bar{\phi}\phi e^{-\frac{1}{3}(K(\Phi,\bar{\Phi}e^{V})+\xi trV)}+\{\int d^{6}z(\phi^{3}W+f_{ab}{\cal W}^{a}{\cal W}^{b}+h.c.\}\label{actiongauge}\end{equation}

The invariance of $K$ gives,\[
(i\Lambda-i\tilde{\Lambda})\frac{\delta K}{\delta V}+\Lambda^{a}k_{a}^{i}K_{i}+\tilde{\Lambda}^{a}k_{a}^{\bar{i}}K_{\bar{i}}=0\]

where $k_{a}$ is the Killing vector corresponding to the gauge generator
with label $a$. From this we have the relation\begin{equation}
\frac{\delta K}{\delta V^{a}}=ik_{a}^{i}K_{i}\label{killing}\end{equation}

Under the gauge transformation the FI term transforms $\xi trV\rightarrow\xi trV+\xi tr(i\Lambda-i\bar{\Lambda})$.
To make the action invariant we need to have the chiral compensator
field also transform under U(1) gauge transformations as $\phi\rightarrow e^{itr\Lambda\frac{\xi}{3}}\phi$.
Then in order to keep the superpotential term invariant we need\[
\delta(\phi^{3}W)=3i\Lambda^{a}trT^{a}\frac{\xi}{3}\phi^{3}W+\phi^{3}\Lambda^{a}k_{a}^{i}\partial_{i}W=0\]

giving us the useful relation \begin{equation}
k_{a}^{i}\partial_{i}W=-i\xi trT^{a}W\label{WFIrelation}\end{equation}

It is convenient to redefine $\phi\rightarrow\phi/W^{1/3}$and eliminate
the superpotential from the action (\ref{actiongauge}) which is now
rewritten with $K\rightarrow G=K+\ln|W|^{2}$ and $W\rightarrow1$.
Then we have the following equations of motion.

\begin{eqnarray*}
-\frac{1}{4}\bar{D}^{2}(\bar{\phi}e^{-\tilde{G}/3}) & = & \phi^{2}\\
-\phi\bar{\phi}e^{-\tilde{G}/3}\frac{1}{4}\bar{\nabla}^{2}G_{i} & = & -\phi^{3}G_{i}-\frac{1}{4}f_{ab,i}(\Phi){\cal W}^{\alpha a}{\cal W}_{\alpha}^{b}\\
-\phi\bar{\phi}e^{-\tilde{G}/3}ik_{a}^{i}G_{i}+\frac{1}{2}\nabla^{\alpha}(f_{ab}{\cal W}_{\alpha}^{b}) & + & \frac{1}{2}\bar{\nabla}^{\dot{\alpha}}(\bar{f}_{ab}{\cal W}_{\dot{\alpha}}^{b})=0\end{eqnarray*}

In the above we have ignored fermionic terms and used (\ref{killing})
with $K\rightarrow G$. Also in getting the the second equation we
used the first. The third equation gives the D-term ($D^{a}=\frac{1}{2}\nabla^{\alpha}{\cal W}_{\alpha}|)$\[
2\Re f_{ab}D^{b}=\phi\bar{\phi}e^{-\tilde{G}/3}|ik^{ai}G_{i}|\]

Choosing the Einstein frame (see discussion after (\ref{einterm}))
$\phi|=\bar{\phi}|=e^{\tilde{G}/6}|=e^{G/6}$| (the last is valid
in WZ gauge) we finally have,\begin{equation}
2\Re f_{ab}D^{b}=ik^{ai}G_{i}|=ik^{ai}\frac{D_{i}W}{W}|=ik^{ai}K_{i}+\xi trT^{a}\label{Dtermfinal}\end{equation}

where in the last equality we've used the relation (\ref{WFIrelation}).
The last form of the D-term is the more familiar one and may be used
whether or not $W$ is zero. However the penultimate form shows that
at minima where the F-terms $D_{i}W$ are zero with ($W\ne0$) i.e.
at a generic supersymmeric AdS minimum, the D-term is also zero. In
other words such a minimum cannot be lifted by D-terms even with an
FI term.

Now we are in a position to calculate the additional contribution
to the potential from the gauge theory D-terms. These are\[
-\int d^{2}\theta\frac{1}{4}f_{ab}{\cal W}^{a\alpha}{\cal W}_{\alpha}^{b}|_{scalar}+h.c.=\frac{1}{2}\Re f_{ab}(\Phi)|D^{a}D^{b}\]

Thus the complete potential for chiral scalars coupled to gauge fields
and supergravity is\begin{equation}
{\cal V}=e^{K}(K^{i\bar{j}}D_{i}WD_{\bar{j}}\bar{W}-3|W|^{2})+\frac{1}{2}\Re f_{ab}(\Phi)|D^{a}D^{b}\label{potfinal}\end{equation}

with the D-term given by (\ref{Dtermfinal}). This is the most important
formula in string (and SUGRA) phenomenology.

\subsection{Quantum effects and the non-perturbative superpotential\label{ssec:VY}}

At the quantum level the Weyl transformations discussed in subsection
(\ref{ssweyl}) are anomalous. The origin of the anomaly is the same
one-loop effect that causes chiral gauge transformations to be anomalous.
In fact the Weyl transformations include an R-transformation. For
putting $\tau(x,\theta)=i\nu$ we see from (\ref{weyl}) that the
chiral fermions transform as \[
\psi_{\alpha}\rightarrow e^{3i\nu}\psi_{\alpha},\,\lambda_{\alpha}\rightarrow e^{-3i\nu}\lambda_{\alpha},\]
where $\psi$ is the fermion in the chiral scalar superfield and $\lambda$
is the gaugino. The anomaly can be thought of as arising from the
non-invariance of the functional integral measure (following Fujikawa's
argument) and by supersymmetrizing the usual (ABJ) expression we have
(note that by abuse of notation we are now labeling the gauge groups
by $a$ rather than the generators of a given group - the trace over
each gauge group is being suppressed)\begin{equation}
[d\Phi][d\bar{\Phi}][dV]\rightarrow e^{i\nu\frac{3c_{a}}{16\pi^{2}}\int d^{8}z\frac{E}{2R}({\cal W}^{\alpha a}{\cal W}_{\alpha}^{a}+h.c.)}[d\Phi][d\bar{\Phi}][dV].\label{measuretrans}\end{equation}

The anomaly coefficient is given by\begin{equation}
c_{a}=T(G_{a})-\sum_{r}n_{r}T_{a}(r)\label{anomcoeff}\end{equation}

where $T_{a}(r)=tr_{r}(T_{a}^{2})$ for a representation $r$ of the
gauge group $G_{a}$, and $r=G_{a}$ means that the corresponding
trace is evaluated in the adjoint representation. Note that by the
standard Adler-Bardeen argument this anomaly is exact.

This anomaly needs to be cancelled since it is actually an anomaly
in a local (super) Weyl symmetry which furthermore is necessary to
obtain Einstein gravity. This can be done by modifying the gauge coupling
function in the following manner \cite{Kaplunovsky:1994fg}: \begin{equation}
f_{a}(\Phi)\rightarrow f_{a}(\Phi,\phi)=f_{a}(\Phi)-\frac{3c_{a}}{8\pi^{2}}\ln\phi.\label{gaugemodify}\end{equation}
The cancellation of the anomalous transformation of the measure then
follows from the Weyl transformation $\phi\rightarrow e^{-2\tau}\phi$
of the Weyl compensator.

If one of the gauge groups that contributes to this anomaly is confining
(exhibits gaugino condensation and develops a mass gap as in QCD)
then this quantum anomaly actually results in a non-perturbative addition
to the classical superpotential \cite{Veneziano:1982ah} (the present
argument is a simplified version of one given in \cite{Burgess:1995aa}).
Non-renormalization theorems imply that the classical superpotential
is not corrected in perturbation theory but there is no such restriction
non-perturbatively.

Suppose the non-perturbatively generated scale of the confining gauge
group is $\Lambda$. Then one expects the gauginos to condense with
$<\lambda\bar{\lambda}>\sim\Lambda^{3}\sim M^{3}e^{-1/g_{YM}^{2}(M)}$
and all fields that are charged under the gauge group to acquire masses
of $O(\Lambda)$. In this case for energies $E\ll\Lambda$ we may
integrate out this gauge theory giving an effective contribution to
the low energy action of the form,\begin{equation}
e^{-\Gamma(\Phi,\phi)}=\int[dV]\exp\left\{ -\frac{1}{4}\int[f_{a}(\Phi)-\frac{3c_{a}}{8\pi^{2}}\ln\phi]{\cal W}^{a\alpha}{\cal W}_{a\alpha}+h.c.\right\} \label{anomeffaction}\end{equation}
(Note that here the chiral fields $\Phi$ are neutral under the group
- fields which are charged are integrated out). However (super) Weyl
invariance tells us that the superpotential term in action must come
multiplied by a factor of $\phi^{3}$. The above then tells us that
the superpotential term in $\Gamma$ must be of the form\[
\phi^{3}W_{np}=w_{a}\phi^{3}e^{-\frac{8\pi^{2}}{c_{a}}f_{a}(\Phi)}.\]

If there are several gauge groups that develop a mass gap then (below
the lowest such scale) there will be a sum of such terms - one for
each gauge group. 

Also since the Kaehler potential in the action occurs with the Weyl
compensator in the combination $\phi\bar{\phi}e^{-K/3}$ the anomaly
canceling term also generates a correction to the Kaehler potential
so that the total Kaehler potential is given by the equation\[
e^{-K/3}=e^{-K_{p}/3}+e^{-K_{np}/3},\]
where $\exp(-K_{np}/3)=k_{a}\exp\{-\frac{8\pi^{2}}{3c_{a}}(f_{a}(\Phi)+\bar{f}_{a}(\bar{\Phi})\}$
and $K_{p}$ is the perturbatively corrected classical Kaehler potential.
Of course since, unlike the superpotential, the Kaehler potential
gets corrected in perturbation theory, these NP terms are less important.

\section{Potential for Moduli in IIB string theory }

\subsection{Classical equations and the flux potential}

We will now discuss the derivation of the effective potential for
the dilaton and the complex structure moduli in type IIB string theory
compactified on Calabi-Yau (CY) orientifolds. We start with the (bosonic
part of the) ten dimensional IIB action (in other words we are in
the supergravity approximation i.e. at energy scales which are much
less than the string scale ($E\ll M_{s}=1/2\pi\alpha'$). The action
is (putting $2\kappa_{10}^{2}=1$)\begin{eqnarray}
S & = & \int d^{10}x\sqrt{g}\{ R-\frac{1}{2\tau_{I}^{2}}\partial_{M}\tau\partial^{M}\bar{\tau}-\frac{1}{2.3!}\frac{1}{\tau_{I}}G_{MNP}\bar{G}^{MNP}\nonumber \\
 & - & \frac{1}{4.5!}\tilde{F}_{MNPQR}\tilde{F}^{MNPQR}\}+\frac{1}{4i}\int\frac{1}{\tau_{I}}C_{4}\wedge G_{3}\wedge\bar{G_{3}}\label{IIBaction}\end{eqnarray}

Here we have put\begin{eqnarray}
\tau=C_{0}+ie^{-\phi}, & G_{3}=F_{3}-\tau H_{3}, & F_{3}=dC_{2},\, H_{3}=dB_{2}\label{tauGdefs}\\
\tilde{F}_{5}=F_{5}-\frac{1}{2}C_{2}\wedge H_{3} & +\frac{1}{2}B_{2}\wedge F_{3}, & F_{5}=dC_{4}\label{F5def}\end{eqnarray}

In the above the numerical index denotes the rank of the corresponding
form, $\phi$ is the dilaton (not to be confused with the field introduced
in the last section) $C_{i}$ is a RR potential of rank $i$ and $B_{2}$
is the NSNS potential.

Now we need to compactify this theory to four dimensions. For phenomenological
reasons it is desirable that the four dimensional theory is ${\cal N}=1$
SUGRA. Since IIB has 32 supersymmetries compactification on a Calabi-Yau
gives eight supersymmetries (i.e. ${\cal N}=2$) in four dimensions
but if we orientifold, we can reduce this to ${\cal N}=1$. However
it will turn out that tadpole cancellations (i.e. satisfaction of
Gauss' law constraints) will require the addition of D-branes and or
fluxes in the internal directions. We will not get into the details
here but it turns out that with a particular orientifold projection
we will end up with orientifold three-planes and D3 and D7 branes
(which actually is a limit of an F-theory construction). For the moment
let us focus on just the three brane action whose leading terms take
the form\begin{equation}
S_{loc}=\sum_{i}(-T_{3}\int_{i}d^{4}x\sqrt{g^{(4)}}+\mu_{3}\int_{i}C_{4})\label{localaction}\end{equation}
This is an action that is localized at a set points $i$ on the CY
orientifold $X$ where the D-branes (orientifolds) are located. The
equations of motion/Bianchi identities coming from the total action
$S+S_{loc}$ are,\begin{eqnarray*}
R_{MN}-\frac{1}{2}g_{MN}R & = & \frac{1}{2}T_{MN}\\
d\frac{*(G_{3}+\bar{G}_{3})}{2\tau_{I}}+H_{3}\wedge\tilde{F}_{5} & = & 0\\
d\frac{*(\tau\bar{G}_{3}+\bar{\tau}G_{3})}{2\tau_{I}}+F_{3}\wedge\tilde{F}_{5} & = & 0\\
d\tilde{F}_{5}=H_{3}\wedge F_{3} & - & \sum_{i}\mu_{3}\delta_{6}^{i}\\
*\tilde{F}_{5} & = & \tilde{F}_{5}\end{eqnarray*}

In the above $\delta_{6}^{i}$ is a delta function on $X$ localized
at the point $i$ where a brane/orientifold plane is located and the
last is the self-duality condition for the five form which is imposed
by hand as usual.

Now we need to do a Kaluza-Klein reduction of these equations to derive
the effective four dimensional equations from these ten dimensional
equations, when as we discussed above six of the dimensions are compactified
on a Calabi-Yau orientifold $X.$ This is unfortunately not as straightforward
as one might think. The problem is that a simple truncation leads
to inconsistencies.

One may try to derive the four dimensional effective action by introducing
the metric ansatz \begin{equation}
ds^{2}=e^{2\omega(y)-6u(x)}\tilde{g}_{\mu\nu}(x)dx^{\mu}dx^{\nu}+e^{-2\omega(y)+2u(x)}\tilde{g}_{mn}(x,y)dy^{m}dy^{n}\label{metric3}\end{equation}

with $\partial_{\mu}\det\tilde{g}_{mn}=0$. The coordinates $x^{\mu}$
are taken over the four dimensional space so that $\tilde{g}_{\mu\nu}$
is the metric measured in our observed world, while $y^{m}$ are the
coordinates on the compact space $X$. For simplicity let us assume
that the number of Kaehler structure moduli in $X$ is just one ($h_{11}=1$)
but we shall keep the number of complex structure moduli $h_{21}$
arbitrary. The Kaehler deformation (which is thus just the volume
modulus corresponding to changes in the overall size of $X$) is given
by the factor $e^{2u}$. In fact we may normalize the internal metric
by putting $\int_{X}e^{-2\omega}\sqrt{\det\tilde{g}_{mn}}$ =1 so
that the physical volume of $X$ is $e^{6u}$. The factor $e^{-6u}$
in the first term is then introduced in order to ensure that $\tilde{g}_{\mu\nu}$
is the four-dimensional Einstein metric. 

The effective potential was derived in \cite{Giddings:2001yu} by
reducing the ten D action using the static version of this ansatz
(i.e with $\partial_{\mu}u=0$ and $\partial_{\mu}g_{mn}=0$ and setting
the warp factor $e^{2\omega(y)}=\alpha$). Thus the expression (the
tilde denotes the use of the metric $\tilde{g}$ in the inner product)
\textcolor{black}{\begin{equation}
V=\int d^{6}y\sqrt{\tilde{g}^{(6)}}\frac{e^{4\omega-12u}}{24\tau_{I}}\widetilde{|iG_{3}-*_{6}G_{3}|^{2}}\label{potential}\end{equation}
}was obtained. However, except at the minimum of the potential the
static ansatz cannot really be used and immediately leads to the no-go
theorem forbidding positive potentials \cite{Gibbons:1984kp}\cite{deWit:1987xg}\cite{Maldacena:2000mw}.
The resolution, as pointed out in \cite{deAlwis:2003sn}, is to include
time dependence of the volume modulus $u(x)$. Furthermore in the
presence of fluxes and D-branes/orientifolds, the warp factor is necessarily
non-trivial, as can be seen from the internal components of the Einstein
equation, and it was shown in \cite{deAlwis:2003sn}
that a consistent derivation was not possible without including all
the Kaluza-Klein (KK) modes, \textit{including
non-diagonal terms in (\ref{metric3})}. In fact it was argued that
the full ten-dimensional equations with time dependent moduli (and
except at the minimum of the potential the moduli are necessarily
time-dependent) and non-trivial warp factor, imply that the metric
ansatz (\ref{metric3}) is invalid. In other words the effective potential
(\ref{potential}) would have additional terms involving KK modes
and the derivatives of the warp factor.

If we ignore these issues (for progress towards resolving them see
\cite{Giddings:2005ff}) then the above potential can be written in
the ${\cal N}=1$ SUGRA form. The superpotential is of the form proposed
in \cite{Gukov:1999ya} (see also \cite{Taylor:1999ii})

\textcolor{black}{\[
W=W_{flux}\equiv\int\Omega\wedge G_{3}.\]
}

\textcolor{black}{Here $\Omega$ is the holomorphic three form on
$X$. $W_{flux}$ depends implicitly on the complex structure moduli
$z_{i}$ (through $\Omega$) and the dilaton $S\equiv i\tau$. However
it is important to note that it is independent of the Kahler modulus
$T$ (with $\Re T=e^{4u}$). Non-renormalization theorems for the
superpotential then imply that this remains true to all orders in
perturbation theory.}

\textcolor{black}{\begin{equation}
K=-\ln(S+\bar{S})-3\ln(T+\bar{T})+k(z^{i},\bar{z}^{\bar{j}}).\label{KahlerSzT}\end{equation}
}

\textcolor{black}{This leads to a no scale potential for $S,\, z_{i}$
\cite{Giddings:2001yu}\[
V=e^{K}(D_{S}WD_{\bar{S}}\bar{W}K^{S\bar{S}}+D_{z}WD_{\bar{z}}\bar{W}K^{z\bar{z}})\ge0.\]
}

\textcolor{black}{This has a minimum at\begin{equation}
D_{S}W=\partial_{S}W+K_{S}W=0,\, D_{z}W=\partial_{z}W+K_{z}W=0\label{Szmin}\end{equation}
}

\textcolor{black}{However generically$F_{T}=D_{T}W_{0}=-3\frac{W_{0}}{T+\bar{T}}\neq0$.
So SUSY is broken in general but the potential at the minimum $V_{0}=0$.
However $T$ ( and so the overall size of the internal manifold) is
unfixed - hence the name no-scale. This is of course unacceptable
since we need to get rid of all Brans-Dicke scalars.}

\subsection{Non-perurbative terms and the KKLT potential}

A proposal to fix \textcolor{black}{$T$ was given by Kachru et al
\cite{Kachru:2003aw} KKLT. Let us discuss this.}

\textcolor{black}{The no-scale feature arises from the fact that the
superpotential is independent of the Kaehler modulus $T$. By standard
non-renormalization theorems the superpotential does not get corrected
in perturbation theory. However there can be non-perturbative terms.
There are two sources for such terms in the present context.}

\textcolor{black}{a) A confining gauge theory on a stack of$D7$ branes
wrapping wrapping a 4-cycle in $X$.}

\textcolor{black}{b) String instanton effects.}

\textcolor{black}{In case a) one can see from examining the DBI action
for a stack of D7-branes that the gauge theory has a gauge coupling
function $f$ (see subsection (\ref{ssec:guage}) that is proportional
to $\Re T$. The latter comes from the volume of the 4-cycle which
is simply given in terms of $\Re T$ since we only have one Kaehler
modulus. It should also be observed that in the Einstein frame there
is no additional modulus dependence of $f$ at the classical level.
By the arguments of subsection (\ref{ssec:VY}) we then get an additional
term in the superpotential so that the total superpotential becomes}

\textcolor{black}{\begin{equation}
W=W_{flux}+Ce^{-aT},\label{Wtotal}\end{equation}
where $C,a$ are constants that depend on the particular D-brane configuration.
A similar term arises from mechanism b). In either case a non-vanishing
pre-factor $C$ requires that the configuration generating it have
two fermionic zero modes corresponding to the $d^{2}\theta$ integration
of a superpotential term. }

\textcolor{black}{KKLT proceed to discuss the physics of the potential
that is generated from this by resorting to a two stage argument.
First ignore the NP term and solve for $S,z$ by imposing (\ref{Szmin}).
This results in a constant superpotential $W_{0}$ which is simply
$W_{flux}$ evaluated at the solution to (\ref{Szmin}). Now add the
NP term. Then we have a potential for the single (complex) modulus
$T$ with the Kahler potential given by the second term of (\ref{KahlerSzT})
(up to a constant) and the superpotential by (\ref{Wtotal}) with $W_{flux}\rightarrow W_{0}.$
The resulting potential just has one AdS SUSY min. with $F_{T}=\partial_{T}W-3\frac{W}{T+\bar{T}}=0$
and $V_{0}=-3|W|^{2}/(T+\bar{T})^{3}$. The self-consistency of the
argument requires that the value at which the stabilization occurs
is such that $\Re T>>1$ $a\Re T>>1$. The first of these comes from
the fact that since our starting point is the ten dimensional supergravity
we are effectively assuming that the Kaluza-Klein mass scale (proportional
to $T^{-1/4}$) is small compared to the string scale. The second
comes from the fact that the NP term is just the leading term in an
expansion (an instanton sum) which is only valid only if this is satisfied.
Then the stabilization condition requires that $W_{0}<<1.$ Generically
of course $W_{0}$ is of order unity, but given a sufficient number
of fluxes it is possible that there would be configurations where
this is satisfied.}

\textcolor{black}{To get a SUSY breaking dS minimum KKLT add a $\bar{D}_{3}$
term to the SUGRA potential to get a potential }

\textcolor{black}{\[
V=V_{sugra}+\frac{d}{(ReT)^{3}}.\]
}

\textcolor{black}{This procedure is rather ad hoc! The Dbar branes
break SUSY at string scale (there would be tower of string states
associated with it which break the opposite half of supersymmetry
to the D-brane/orientifold system). Thus it is unclear how one can
use the four dimensional $N=1$ SUGRA formalism at all. A rigorous
derivation would require that we try to repeat the arguments of GKP
in the presence of the Dbar branes. Given that we already had problems
of decoupling KK modes even in the case with more control it seems
unlikely that one could find a rigorous derivation of such a potential
with SUSY broken already at the ten-dimensional level. In fact even
if the naive truncation is done we would have ended up with a runaway
potential for $T$ at the first stage of the KKLT argument!}

\textcolor{black}{Can the Dbar term be interpreted as a D term in
N=1 SUGRA? This is unclear but even if this is true, there is no uplift
of a F-term supersymmeric AdS vacuum. As we discussed in subsection
(\ref{ssec:guage}) the F and D terms are related (when the superpotential
is non-vanishing) by}

\textcolor{black}{\[
f_{ab}D^{b}=\frac{ik_{a}^{i}D_{i}W}{W}\]
}

\textcolor{black}{So for an F-term supersymmetric minimum which is
AdS i.e. $D_{i}W=0,\, W\ne0$, the D-term also vanishes so that it
cannot be used to uplift the potential }%
\footnote{Burgess et al give a proposal for getting explicit FI D terms\cite{Burgess:2003ic}.
The above argument shows that they cannot be used to lift AdS SUSY
minima as in the KKLT case but some such mechanism might work if the
F term potential gave a SUSY breaking AdS minimum.%
}\textcolor{black}{. One might ask however whether by including more
NP terms in the superpotential one can get a dS minimum by F-terms
alone. Within the context of the two stage procedure this is impossible
though it can be done if at least two light moduli are included\cite{Brustein:2004xn}.}

\textcolor{black}{However one should ask whether this two step procedure
is justified - even assuming large masses for $S,$$z$? The full
superpotential is given by (\ref{Wtotal}) and it is clear that even
if one is justified in assuming that for some range of values of $T$
the masses of $S$ and $z$ are heavy it is clear that the equations
for integrating out these fields will not set them to be constant
but to be functions of $T$. Thus the correct procedure would yield
a potential that is much more complicated than what one obtains in
the two stage calculation where the {}``heavy'' fields have been
set to constants. There is in fact no approximation scheme in which
the two stage procedure is justified. However we will find that assuming
there are flux configurations such that $M_{S},M_{z}\sim M_{KK}$
so that we can integrate out $S,z$, it is possible to avoid the ad
hoc uplifting procedure. }

\subsection{Integrating out in SUSY and SUGRA}

Before we discuss the KKLT potential however it is worthwhile discussing
the general procedure of (classically) integrating out fields in field
theory. \textcolor{black}{There does not seem to be much discussion
of these issues in the case of supersymmetric (especially supergravity)
theories so it is perhaps worthwhile going through this exercise in
some detail.}

\textcolor{black}{Suppose the potential of a scalar field theory with
a heavy ($\Phi$) and a light ($\phi$) field is\[
V(\Phi,\phi)=\frac{1}{2}M^{2}\Phi^{2}+\tilde{V}(\Phi,\phi)\]
 Solving the equation of motion (EOM) for $\Phi$ we get \[
\Phi=\frac{1}{\square-M^{2}}\frac{\partial\tilde{V}}{\partial\Phi}=-\frac{1}{M^{2}}\frac{\partial\tilde{V}}{\partial\Phi}+\frac{1}{M^{2}}O(\frac{\square}{M^{2}}\frac{\partial\tilde{V}}{\partial\Phi}).\]
}

\textcolor{black}{So up to terms $O(E/M)$ $\Phi$ solves}

\textcolor{black}{\[
\frac{\partial V}{\partial\Phi}=0\]
}

\textcolor{black}{The effective potential for light fields is then
\[
V(\Phi(\phi),\phi).\]
}

\textcolor{black}{Now let us look at a SUSY (global) theory:}

\textcolor{black}{\[
S=\int d^{4}xd^{4}\theta\bar{\Phi^{i}}\Phi^{i}+(\int d^{4}xd^{2}\theta W(\Phi^{i})+c.c.).\]
}

\textcolor{black}{with superfield EOM}

\textcolor{black}{\[
-\frac{1}{4}\bar{D}^{2}\bar{\Phi^{i}}+\frac{\partial W}{\partial\Phi^{i}}=0.\]
Let $i=H$ for heavy and $i=l,l'$ for light fields. The standard
argument for integrating out $\Phi^{H}$ is that one should use \begin{equation}
\frac{\partial W}{\partial\Phi^{H}}=0.\label{SUSYcond1}\end{equation}
 However this condition is valid only under certain restrictions.
Consider the following superpotential }

\textcolor{black}{\[
W(\Phi^{H},\Phi^{l})=\frac{1}{2}M\Phi^{H2}+\tilde{W}(\Phi^{H},\Phi^{l}).\]
}

\textcolor{black}{Solving the E of M for heavy field:}

\textcolor{black}{\[
\Phi^{H}=\frac{1}{\square-M^{2}}(M\frac{\partial\tilde{W}}{\partial\Phi^{H}}+\frac{\bar{D}^{2}}{4}\frac{\partial\bar{\tilde{W}}}{\partial\bar{\Phi}^{H}})\]
}

\textcolor{black}{Expand in powers of $\square/M^{2}$ as before to
get,}

\textcolor{black}{\begin{equation}
\frac{\partial W}{\partial\Phi^{H}}=-\frac{\bar{D}^{2}}{4M}\frac{\partial\bar{\tilde{W}}}{\partial\bar{\Phi}^{H}}+O(\frac{\square}{M^{2}}).\label{SUSYcond2}\end{equation}
}

\textcolor{black}{So we see that when $\tilde{W}\ne0$, to get the
usual condition (\ref{SUSYcond1}) we need in addition to $\frac{E^{2}}{M^{2}}<<1$,
also $|\Phi_{l}|<<M$.}

\textcolor{black}{Let us illustrate this with a simple example having
one heavy $H$ and one light field $L$.\[
\int d^{4}\theta(\bar{H}H+\bar{L}L)+\left[\int d^{2}\theta\frac{1}{2}(MH^{2}+HL^{2})+c.c.\right]\]
}

The potential is easily worked out giving,

\textcolor{black}{\begin{eqnarray*}
-V & = & \bar{F}F+\bar{f}f+\frac{1}{2}(Fa^{2}+\bar{F}\bar{a}^{2})+(Aaf+\bar{A}\bar{a}\bar{f})\\
 & + & M(AF+\bar{A}\bar{F})\end{eqnarray*}
}

\textcolor{black}{Ignoring fermions let us put $H=(A,F)$, $l=(a,f)$.}

\textcolor{black}{The heavy eqns. are\begin{eqnarray*}
-\square\bar{A} & = & MF+af\\
\bar{F} & = & -\frac{a^{2}}{2}-MA\end{eqnarray*}
Plug these solution for $F$ into V ignoring $O(\square/M^{2})$ to
get}

\[
-V=\bar{f}f(1+\frac{\bar{a}a}{M^{2}})-\frac{1}{2M}(fa^{3}+\bar{f}\bar{a}^{3}).\]

\textcolor{black}{Eliminating the light auxiliary field using its
equation of motion,}

\textcolor{black}{\[
V=\frac{|a|^{6}}{4M^{2}}(1+\frac{|a|^{2}}{M^{2}})^{-1}.\]
}

\textcolor{black}{If $\partial W/\partial H=0$ had been used the
term $|a|^{2}/M^{2}$ would not be obtained.}

\textcolor{black}{So the usual condition is valid for }

\textcolor{black}{\[
\frac{|a|^{2}}{M^{2}}<<1.\]
What is the corresponding condition in SUGRA?}

\textcolor{black}{We expect (\ref{SUSYcond1}) to be replaced by}

\textcolor{black}{\begin{equation}
D_{H}W=\partial_{H}W+\frac{1}{M_{p}^{2}}W\partial_{H}K=0.\label{SUGRAcond1}\end{equation}
}

\textcolor{black}{Using the flat space chiral compensator formalism
of GGRS \cite{Gates:1983nr} we have (with $M_{p}^{2}=1)$}

\textcolor{black}{\begin{equation}
S=-3\int d^{4}xd^{4}\theta\bar{\phi}\phi e^{-K/3}+(\int d^{4}xd^{2}\theta\phi^{3}W+h.c.)\label{flatSUGRAaction}\end{equation}
}

\textcolor{black}{Take \begin{equation}
K=\overline{H}H+K^{l}(L,\overline{L}),\label{KHL}\end{equation}
}

\textcolor{black}{with a superpotential \begin{equation}
W=\frac{1}{2}MH^{2}+\tilde{W}(H,L),\label{WHL}\end{equation}
We need to solve heavy field equation ignoring $O(E^{2}/M^{2})$ terms.
Effectively this amounts to putting $\square H=0$. So we get}

\textcolor{black}{\begin{eqnarray}
D^{2}(e^{K/3M_{p}^{2}}\phi^{2})D_{H}W+ & 4Me^{2K/M_{p}^{2}}\phi^{2}\bar{\phi}^{2}(1+\frac{\bar{H}H}{M_{p}^{2}})D_{\bar{H}}\bar{W}\nonumber \\
 & =-e^{K/3M_{p}^{2}}\phi^{2}D^{2}D_{H}\tilde{W}\nonumber \\
 & +O(D_{\alpha}H)\label{SUGRAcond2}\end{eqnarray}
}

\textcolor{black}{Note that the global limit $M_{p}\rightarrow\infty$,
$\phi\rightarrow1$ gives the previous result (\ref{SUSYcond2}).
As in that case here too we can ignore the first term in the RHS for
$|L|<<M$. But what about the fermionic terms? It turns out that $D_{H}W=0$
is a sufficient condition. Spinor derivation of this condition
gives $W\overline{D}_{\dot{\alpha}}\overline{H}$=0 and so at generic
points in field space the fermion terms vanish. However the condition}
$D_{H}W=0$ is not a necessary one - in fact it is too strong \textcolor{black}{and
we cannot ignore the fermion terms (even for $L<<M$). To see this
take\[
K=K^{h}(H,\overline{H})+K^{l}(L,\overline{L})\]
}

\textcolor{black}{The condition (\ref{SUGRAcond1}) is\begin{equation}
\partial_{H}W+\partial_{H}K^{h}W=0\label{eq:}\end{equation}
}

\textcolor{black}{This is not a chiral eqn. Take the anti-chiral derivative
to get}

\textcolor{black}{\begin{equation}
WK_{H\bar{H}}^{h}{\cal \overline{\mathcal{D}}_{\dot{\alpha}}}\bar{H}=0\label{eq2:}\end{equation}
}

\textcolor{black}{Suppose (\ref{eq:}) is solved by $H=H(L,\bar{L}).$
From chirality of $H,L$ and (\ref{eq2:}),}

\textcolor{black}{\[
\bar{\mathcal{D}}_{\dot{\alpha}}H=\frac{\partial H}{\partial L}\bar{\mathcal{D}}_{\dot{\alpha}}L+\frac{\partial H}{\partial\bar{L}}\bar{\mathcal{D}}_{\dot{\alpha}}\bar{L}=\frac{\partial H}{\partial\bar{L}}\bar{\mathcal{D}}_{\dot{\alpha}}\bar{L}=0.\]
}

\textcolor{black}{\[
\mathcal{D}_{\alpha}H=\frac{\partial H}{\partial L}\mathcal{D}_{\alpha}L+\frac{\partial H}{\partial\bar{L}}\mathcal{D}_{\alpha}\bar{L}=\frac{\partial H}{\partial L}\mathcal{D}_{\alpha}L=0\]
}

\textcolor{black}{This implies that there are no fermions in the light
field theory! The lesson is that $D_{H}W=0$ can be imposed for computing
scalar potential (for $L<<M$) but should not be used as a superfield
relation, since in that case we would need to keep the fermion squared
terms in (\ref{SUGRAcond2}).}

\textcolor{black}{Note that even with $L<<M<M_{p}$, non-trivial SUGRA
terms will be present since }

\textcolor{black}{\[
W(L.\bar{L)}=W_{0}+....\]
with $W_{0}\sim O(M_{p})$ in typical situations so SUGRA corrections
to global SUSY $K_{L}W/M_{p}^{2},\,\,|W|^{2}/M_{p}^{2}$ are not necessarily
negligible!}

\subsection{String theory potentials for light moduli}

\textcolor{black}{Let us now get back to string theory effective potentials.
First consider a simple model where the compact dimensions are taken
to be on a rigid CY manifold. This means we just have the Kaehler
modulus that corresponds to changing the overall size of the manifold.
This then gives a model with just $S$ and $T$.}

\textcolor{black}{The classical Kaehler potential is}

\textcolor{black}{\[
K=-\ln(S+\bar{S)}-3\ln(T+\bar{T})\]
}

\textcolor{black}{The superpotential coming from fluxes and a NP term
is }

\textcolor{black}{\[
W=A+SB+Ce^{-aT}\]
}

\textcolor{black}{Assume that the fluxes are such that $S$ is heavy
so we may integrate it out using}

\textcolor{black}{\[
D_{S}W=B-\frac{A+SB+Ce^{-aT}}{S+\bar{S}}=0.\]
}

\textcolor{black}{This is solved by}

\textcolor{black}{\[
\bar{S}=(A+Ce^{-aT})/B\]
}

\textcolor{black}{This equation is not holomorphic! This is not necessarily
a problem since the SUGRA action is determined by the Kaehler invariant
combination }

\textcolor{black}{\[
G=K(\Phi,\bar{\Phi})+\ln W(\Phi)+\ln\bar{W}(\bar{\Phi}).\]
}

\textcolor{black}{After using the solution for $S$ we get an effective
potential for $T$ that is determined by}

\textcolor{black}{\begin{eqnarray*}
G & = & -\ln(\frac{(A+Ce^{-aT})}{B}+\frac{(\bar{A}+\bar{C}e^{-a\bar{T}})}{\bar{B}})-3\ln(T+\bar{T)}\\
 &  & +\ln(A+B\frac{(\bar{A}+\bar{C}e^{-a\bar{T}})}{\bar{B}}+Ce^{-aT})+c.c.).\end{eqnarray*}
}

\textcolor{black}{On the other hand if the KKLT two stage process
is used, $S=A/B$, so that}

\textcolor{black}{\[
G=-3\ln(T+\bar{T})+(\ln(A+B\frac{\bar{A}}{\bar{B}}+Ce^{-aT})+c.c.)\]
}

\textcolor{black}{The problem is that the missing NP terms, for example
the term }

\textcolor{black}{\[
\frac{B}{\bar{B}}\bar{C}e^{-a\bar{T}}\]
is of the same magnitude as the ones that are being kept! Note that
this calculation would still be valid when there is more than one
Kaehler modulus (but keeping $h_{21}=0$) with the replacement $Ce^{-aT}\rightarrow\sum_{i}C_{i}e^{-a_{i}T_{i}}$.}

\textcolor{black}{This model actually has no minima in $S,T$ space
as shown in \cite{Choi:2004sx}. The SUSY extremum at $D_{S}W=D_{T}W$=0
is a saddle point. Of course since we have AdS supersymmetry that's
OK! But obviously it cannot even be uplifted by $\bar{D}$ terms to
get a phenomenologically viablemodel. }

\textcolor{black}{Let us now consider models with complex structure
moduli $z^{i}$ but with just one Kaehler modulus:\[
K=-\ln(S+\bar{S})-3\ln(T+\bar{T})+k(z^{i},\bar{z}^{\bar{j}})\]
}

\textcolor{black}{\[
W=A(z^{i})+SB(z^{i})+Ce^{-aT}.\]
As before, the first two terms in $W$ give the flux superpotential
and the third term is coming from NP effects. The Kaehler derivatives
are \begin{eqnarray*}
D_{T}W & = & -aCe^{-aT}-\frac{3}{T+\bar{T}}W,\\
D_{S}W & = & B-\frac{W}{S+\bar{S}},\\
D_{i}W & = & \partial_{i}A+S\partial_{i}B+\partial_{i}kW.\end{eqnarray*}
Let us assume that the $z^{i}$ are heavy and integrate them out using
$D_{i}W=0$. Now unlike in the previous case we cannot solve this
explicitly but it is easy to show that there is no holomorphic solution
contrary to claims in the literature.}

\textcolor{black}{Suppose now that there is a holomorphic solution.
So}

\textcolor{black}{\[
W=W_{eff}+Ce^{-aT},\]
}

\textcolor{black}{with\[
W_{eff}=A(z^{i}(S,T))+B(z^{i}(S,T))S,\]
}

\textcolor{black}{and\[
K=-\ln(S+\bar{S})-3\ln(T+\bar{T})+k(z(S,T),\bar{z}(\bar{S},\bar{T})).\]
}

\textcolor{black}{The eqn to be solved is }

\textcolor{black}{\begin{equation}
D_{i}W=\partial_{i}A(z^{i})+S\partial_{i}B(z^{i})+W(S,T,z^{i})k_{i}=0.\label{zcond}\end{equation}
}

\textcolor{black}{The assumption is that this is solved by $z^{i}=z^{i}(S,T)$.
Differentiate (\ref{zcond}) w.r.t. $\bar{S}$;}

\textcolor{black}{\[
W(S,T,z^{i}(S,T))k_{i,\bar{j}}\frac{\partial\bar{z}^{\bar{j}}}{\partial\bar{S}}=0.\]
}

\textcolor{black}{At generic points $W\ne0$ and $k_{i\bar{j}}$ is
non-degenerate.}

\textcolor{black}{So \[
\frac{\partial z^{i}}{\partial S}=0\]
}

\textcolor{black}{Similarly we get $\frac{\partial z^{i}}{\partial T}=0$.
But this obviously cannot be the case so we conclude that the $z^{i}$
cannot be holomorphic functions of $S,$T. So we expect a solution
of the form $z^{i}=z^{i}(S,T,\bar{S,}\bar{T})$.}

\textcolor{black}{As in the simple case (with no $z$) that we solved
explicitly, we need to compute V from\begin{eqnarray*}
G=K & = & -\ln(S+\bar{S})-3\ln(T+\bar{T})\\
 & + & k(z(S,T,\bar{S},\bar{T}),\bar{z}(\bar{S},\bar{T},S,T))\\
 & + & \ln|W(S,T,z(S,T,\bar{S},\bar{T}))|^{2}\end{eqnarray*}
}

\textcolor{black}{Since it is hard to deal with two complex variables
let us assume that $S$ is heavy as well for some choice of fluxes.
i.e put $D_{S}W=0$ so that\[
z^{i}=z^{i}(T,\bar{T})\,{\rm and}\, S=S(T,\bar{T}).\]
Let us assume that there is a power series solution, valid for $a\Re T>>1$. }

\textcolor{black}{\begin{eqnarray*}
S & = & \alpha+\beta Ce^{-aT}+\gamma\bar{C}e^{-a\bar{T}}+...\\
z^{i} & = & \alpha^{i}+\beta^{i}Ce^{-aT}+\gamma^{i}\bar{C}e^{-a\bar{T}}+...\end{eqnarray*}
}

\textcolor{black}{where $\alpha,...\gamma^{i}$ are functions of flux
integers. Then}

\textcolor{black}{\begin{eqnarray*}
G & = & \ln(v+bCe^{-aT}+\bar{b}\bar{C}e^{-a\bar{T}}+cC^{2}e^{-2aT}+\bar{c}\bar{C}^{2}e^{-2a\bar{T}}\\
 &  & +d|C|^{2}e^{-a(T+\bar{T})}+...)-3\ln(T+\bar{T}),\end{eqnarray*}
}

\textcolor{black}{and the potential becomes}

\textcolor{black}{\begin{eqnarray*}
{\cal V} & = & \frac{1}{(T+\bar{T})^{2}}[a(bCe^{-aT}+2cC^{2}e^{-2aT}+c.c.)\\
 & + & a|C|^{2}((4\frac{a|b|^{2}}{v}-3ad)\frac{T+\bar{T}}{3}+2d)e^{-a(T+\bar{T})}].\end{eqnarray*}
}

\textcolor{black}{This differs from a two stage calculation by terms
such as }

\textcolor{black}{\[
2cC^{2}e^{-2aT}+c.c.\]
}

\textcolor{black}{which are obviously of the same magnitude as terms
which are kept. Note that even in the real direction this potential
has more parameters than in the two stage process. The reason is that
in the latter all the flux parameters are subsumed into $W_{0}$.}

\textcolor{black}{With the two stage version we can show that there
are no dS minima even with multiple condensate terms.\cite{Brustein:2004xn}.
The above on the other hand can have positive minima!}

An explicit example is the following

\includegraphics{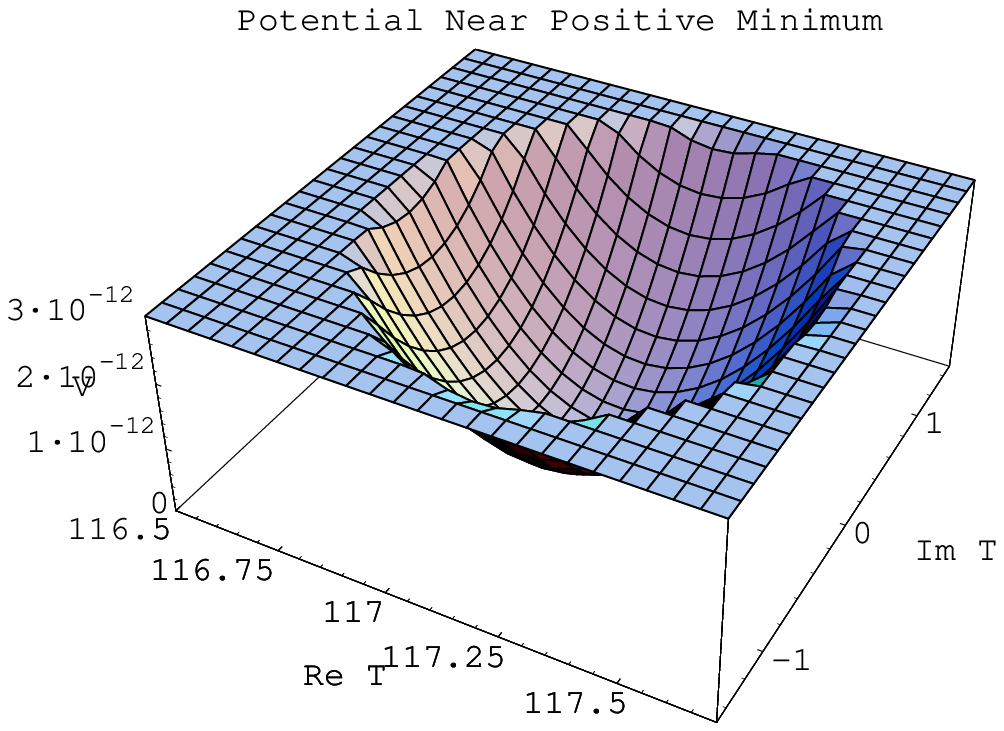}

We choose the following parameters: \begin{eqnarray*}
a=\frac{2\pi}{320} & v=0.22941751641574312 & b=1\\
 & c=-1.4097828718993035\\
 & d=15.786002156414208\end{eqnarray*}

This then has a local minimum at $\Re T_{min}=117.138,\,\Im T=0,\, V_{min}=10^{-15}$.
\textcolor{black}{The example is of course somewhat artificial and
certainly does not give a viable phenomenology. What it does demonstrate
is that the uplift terms of KKLT are not really essential. Finding
a viable model without them however might be a challenge! Of course
in any case it is not really necessary to get dS minima at the classical
level - quantum contributions from integrating out high frequency
modes from the string scale down to cosmological scales, together
with contributions from the standard model and QCD phase transitions
could well lift a ADS vacuum whose CC is not too large (i.e. of the
order of the standard model scale).}

\subsection{\textcolor{black}{Conclusions}}

\textcolor{black}{In this section we argued that in SUSY theories
imposing $\partial_{H}W=0$ to integrate out a heavy field $H$ is
valid if the light field space is restricted to $|L|<<M$. In SUGRA
the corresponding equation is $D_{H}W=0$, but it has to be used with
caution - in particular it is not valid as an equation for superfields,
but it is valid subject to the same restrictions as in the SUSY case,
for the purpose of calculating the potential.}

In applying this to potentials arising from flux compactifications
corrected by non-perturbative terms, in order to derive an effective
potential for a light Kaehler modulus we found that the correct procedure
leads to a more complicated potential than what one gets if one followed
the two stage procedure of KKLT. The additional terms are of the same
magnitude as the terms which kept in the two stage procedure. It was
argued that these may lead to phenomenologically viable models for
inflation without the need for ad hoc uplifting terms.

\section{Acknowledgements}

\textcolor{black}{I wish to thank Shamit Kachru and Eva Silverstein
for inviting me to give these lectures at TASI 2005, JianXin Lu for
inviting me to lecture at the Shanghai Summer School on M-theory 2005,
and Rob Myers for inviting me to give these lectures at the Perimeter
Institute.}

\textcolor{black}{I also wich to thank Ramy Brustein, Marc Grisaru,
Peter Nilles, and especially Martin Rocek for discussions on the content
of section 3 and Paul Martens for the Mathematica plot displayed there.}

This \textcolor{black}{work is supported by DOE grant No. DE-FG02-91-ER-40672.}

\bibliographystyle{apsrev}
\bibliography{myrefs}

\begin{thebibliography}{32}
\expandafter\ifx\csname natexlab\endcsname\relax\def\natexlab#1{#1}\fi
\expandafter\ifx\csname bibnamefont\endcsname\relax
  \def\bibnamefont#1{#1}\fi
\expandafter\ifx\csname bibfnamefont\endcsname\relax
  \def\bibfnamefont#1{#1}\fi
\expandafter\ifx\csname citenamefont\endcsname\relax
  \def\citenamefont#1{#1}\fi
\expandafter\ifx\csname url\endcsname\relax
  \def\url#1{\texttt{#1}}\fi
\expandafter\ifx\csname urlprefix\endcsname\relax\def\urlprefix{URL }\fi
\providecommand{\bibinfo}[2]{#2}
\providecommand{\eprint}[2][]{\url{#2}}

\bibitem[{\citenamefont{Wess and Bagger}(1992)}]{Wess:1992cp}
\bibinfo{author}{\bibfnamefont{J.}~\bibnamefont{Wess}} \bibnamefont{and}
  \bibinfo{author}{\bibfnamefont{J.}~\bibnamefont{Bagger}},
  \bibinfo{journal}{Supersymmetry and supergravity}  (\bibinfo{year}{1992}),
  \bibinfo{note}{princeton, USA: Univ. Pr. 259 p}.

\bibitem[{\citenamefont{Gates et~al.}(1983)\citenamefont{Gates, Grisaru, Rocek,
  and Siegel}}]{Gates:1983nr}
\bibinfo{author}{\bibfnamefont{S.~J.} \bibnamefont{Gates}},
  \bibinfo{author}{\bibfnamefont{M.~T.} \bibnamefont{Grisaru}},
  \bibinfo{author}{\bibfnamefont{M.}~\bibnamefont{Rocek}}, \bibnamefont{and}
  \bibinfo{author}{\bibfnamefont{W.}~\bibnamefont{Siegel}},
  \bibinfo{journal}{Superspace, or one thousand and one lessons in
  supersymmetry - Front. Phys.} \textbf{\bibinfo{volume}{58}},
  \bibinfo{pages}{1} (\bibinfo{year}{1983}), \eprint{hep-th/0108200}.

\bibitem[{\citenamefont{Kaplunovsky and Louis}(1994)}]{Kaplunovsky:1994fg}
\bibinfo{author}{\bibfnamefont{V.}~\bibnamefont{Kaplunovsky}} \bibnamefont{and}
  \bibinfo{author}{\bibfnamefont{J.}~\bibnamefont{Louis}},
  \bibinfo{journal}{Nucl. Phys.} \textbf{\bibinfo{volume}{B422}},
  \bibinfo{pages}{57} (\bibinfo{year}{1994}), \eprint{hep-th/9402005}.

\bibitem[{\citenamefont{Burgess et~al.}(1996)\citenamefont{Burgess,
  Derendinger, Quevedo, and Quiros}}]{Burgess:1995aa}
\bibinfo{author}{\bibfnamefont{C.~P.} \bibnamefont{Burgess}},
  \bibinfo{author}{\bibfnamefont{J.~P.} \bibnamefont{Derendinger}},
  \bibinfo{author}{\bibfnamefont{F.}~\bibnamefont{Quevedo}}, \bibnamefont{and}
  \bibinfo{author}{\bibfnamefont{M.}~\bibnamefont{Quiros}},
  \bibinfo{journal}{Annals Phys.} \textbf{\bibinfo{volume}{250}},
  \bibinfo{pages}{193} (\bibinfo{year}{1996}), \eprint{hep-th/9505171}.

\bibitem[{\citenamefont{Giddings et~al.}(2002)\citenamefont{Giddings, Kachru,
  and Polchinski}}]{Giddings:2001yu}
\bibinfo{author}{\bibfnamefont{S.~B.} \bibnamefont{Giddings}},
  \bibinfo{author}{\bibfnamefont{S.}~\bibnamefont{Kachru}}, \bibnamefont{and}
  \bibinfo{author}{\bibfnamefont{J.}~\bibnamefont{Polchinski}},
  \bibinfo{journal}{Phys. Rev.} \textbf{\bibinfo{volume}{D66}},
  \bibinfo{pages}{106006} (\bibinfo{year}{2002}), \eprint{hep-th/0105097}.

\bibitem[{\citenamefont{Kachru et~al.}(2003)\citenamefont{Kachru, Kallosh,
  Linde, and Trivedi}}]{Kachru:2003aw}
\bibinfo{author}{\bibfnamefont{S.}~\bibnamefont{Kachru}},
  \bibinfo{author}{\bibfnamefont{R.}~\bibnamefont{Kallosh}},
  \bibinfo{author}{\bibfnamefont{A.}~\bibnamefont{Linde}}, \bibnamefont{and}
  \bibinfo{author}{\bibfnamefont{S.~P.} \bibnamefont{Trivedi}}
  (\bibinfo{year}{2003}), \eprint{hep-th/0301240}.

\bibitem[{\citenamefont{Dine et~al.}(1985)\citenamefont{Dine, Rohm, Seiberg,
  and Witten}}]{Dine:1985rz}
\bibinfo{author}{\bibfnamefont{M.}~\bibnamefont{Dine}},
  \bibinfo{author}{\bibfnamefont{R.}~\bibnamefont{Rohm}},
  \bibinfo{author}{\bibfnamefont{N.}~\bibnamefont{Seiberg}}, \bibnamefont{and}
  \bibinfo{author}{\bibfnamefont{E.}~\bibnamefont{Witten}},
  \bibinfo{journal}{Phys. Lett.} \textbf{\bibinfo{volume}{B156}},
  \bibinfo{pages}{55} (\bibinfo{year}{1985}).

\bibitem[{\citenamefont{Strominger}(1986)}]{Strominger:1986uh}
\bibinfo{author}{\bibfnamefont{A.}~\bibnamefont{Strominger}},
  \bibinfo{journal}{Nucl. Phys.} \textbf{\bibinfo{volume}{B274}},
  \bibinfo{pages}{253} (\bibinfo{year}{1986}).

\bibitem[{\citenamefont{Polchinski and Strominger}(1996)}]{Polchinski:1995sm}
\bibinfo{author}{\bibfnamefont{J.}~\bibnamefont{Polchinski}} \bibnamefont{and}
  \bibinfo{author}{\bibfnamefont{A.}~\bibnamefont{Strominger}},
  \bibinfo{journal}{Phys. Lett.} \textbf{\bibinfo{volume}{B388}},
  \bibinfo{pages}{736} (\bibinfo{year}{1996}), \eprint{hep-th/9510227}.

\bibitem[{\citenamefont{Becker and Becker}(1996)}]{Becker:1996gj}
\bibinfo{author}{\bibfnamefont{K.}~\bibnamefont{Becker}} \bibnamefont{and}
  \bibinfo{author}{\bibfnamefont{M.}~\bibnamefont{Becker}},
  \bibinfo{journal}{Nucl. Phys.} \textbf{\bibinfo{volume}{B477}},
  \bibinfo{pages}{155} (\bibinfo{year}{1996}),
  \eprint[http://arXiv.org/abs]{hep-th/9605053}.

\bibitem[{\citenamefont{Dasgupta et~al.}(1999)\citenamefont{Dasgupta, Rajesh,
  and Sethi}}]{Dasgupta:1999ss}
\bibinfo{author}{\bibfnamefont{K.}~\bibnamefont{Dasgupta}},
  \bibinfo{author}{\bibfnamefont{G.}~\bibnamefont{Rajesh}}, \bibnamefont{and}
  \bibinfo{author}{\bibfnamefont{S.}~\bibnamefont{Sethi}},
  \bibinfo{journal}{JHEP} \textbf{\bibinfo{volume}{08}}, \bibinfo{pages}{023}
  (\bibinfo{year}{1999}), \eprint{hep-th/9908088}.

\bibitem[{\citenamefont{Taylor and Vafa}(2000)}]{Taylor:1999ii}
\bibinfo{author}{\bibfnamefont{T.~R.} \bibnamefont{Taylor}} \bibnamefont{and}
  \bibinfo{author}{\bibfnamefont{C.}~\bibnamefont{Vafa}},
  \bibinfo{journal}{Phys. Lett.} \textbf{\bibinfo{volume}{B474}},
  \bibinfo{pages}{130} (\bibinfo{year}{2000}),
  \eprint[http://arXiv.org/abs]{hep-th/9912152}.

\bibitem[{\citenamefont{Gukov et~al.}(2000)\citenamefont{Gukov, Vafa, and
  Witten}}]{Gukov:1999ya}
\bibinfo{author}{\bibfnamefont{S.}~\bibnamefont{Gukov}},
  \bibinfo{author}{\bibfnamefont{C.}~\bibnamefont{Vafa}}, \bibnamefont{and}
  \bibinfo{author}{\bibfnamefont{E.}~\bibnamefont{Witten}},
  \bibinfo{journal}{Nucl. Phys.} \textbf{\bibinfo{volume}{B584}},
  \bibinfo{pages}{69} (\bibinfo{year}{2000}), \eprint{hep-th/9906070}.

\bibitem[{\citenamefont{Greene et~al.}(2000)\citenamefont{Greene, Schalm, and
  Shiu}}]{Greene:2000gh}
\bibinfo{author}{\bibfnamefont{B.~R.} \bibnamefont{Greene}},
  \bibinfo{author}{\bibfnamefont{K.}~\bibnamefont{Schalm}}, \bibnamefont{and}
  \bibinfo{author}{\bibfnamefont{G.}~\bibnamefont{Shiu}},
  \bibinfo{journal}{Nucl. Phys.} \textbf{\bibinfo{volume}{B584}},
  \bibinfo{pages}{480} (\bibinfo{year}{2000}), \eprint{hep-th/0004103}.

\bibitem[{\citenamefont{Grana and Polchinski}(2001)}]{Grana:2000jj}
\bibinfo{author}{\bibfnamefont{M.}~\bibnamefont{Grana}} \bibnamefont{and}
  \bibinfo{author}{\bibfnamefont{J.}~\bibnamefont{Polchinski}},
  \bibinfo{journal}{Phys. Rev.} \textbf{\bibinfo{volume}{D63}},
  \bibinfo{pages}{026001} (\bibinfo{year}{2001}), \eprint{hep-th/0009211}.

\bibitem[{\citenamefont{de~Alwis}(2003)}]{deAlwis:2003sn}
\bibinfo{author}{\bibfnamefont{S.~P.} \bibnamefont{de~Alwis}},
  \bibinfo{journal}{Phys. Rev.} \textbf{\bibinfo{volume}{D68}},
  \bibinfo{pages}{126001} (\bibinfo{year}{2003}), \eprint{hep-th/0307084}.

\bibitem[{\citenamefont{de~Alwis}(2004)}]{deAlwis:2004qh}
\bibinfo{author}{\bibfnamefont{S.~P.} \bibnamefont{de~Alwis}},
  \bibinfo{journal}{Phys. Lett.} \textbf{\bibinfo{volume}{B603}},
  \bibinfo{pages}{230} (\bibinfo{year}{2004}), \eprint{hep-th/0407126}.

\bibitem[{\citenamefont{de~Alwis}(2005{\natexlab{a}})}]{deAlwis:2005tg}
\bibinfo{author}{\bibfnamefont{S.~P.} \bibnamefont{de~Alwis}},
  \bibinfo{journal}{Phys. Lett.} \textbf{\bibinfo{volume}{B626}},
  \bibinfo{pages}{223} (\bibinfo{year}{2005}{\natexlab{a}}),
  \eprint{hep-th/0506267}.

\bibitem[{\citenamefont{de~Alwis}(2005{\natexlab{b}})}]{deAlwis:2005tf}
\bibinfo{author}{\bibfnamefont{S.~P.} \bibnamefont{de~Alwis}},
  \bibinfo{journal}{Phys. Lett.} \textbf{\bibinfo{volume}{B628}},
  \bibinfo{pages}{183} (\bibinfo{year}{2005}{\natexlab{b}}),
  \eprint{hep-th/0506266}.

\bibitem[{\citenamefont{Frey}(2003)}]{Frey:2003tf}
\bibinfo{author}{\bibfnamefont{A.~R.} \bibnamefont{Frey}}
  (\bibinfo{year}{2003}), \eprint{hep-th/0308156}.

\bibitem[{\citenamefont{Balasubramanian}(2004)}]{Balasubramanian:2004wx}
\bibinfo{author}{\bibfnamefont{V.}~\bibnamefont{Balasubramanian}},
  \bibinfo{journal}{Class. Quant. Grav.} \textbf{\bibinfo{volume}{21}},
  \bibinfo{pages}{S1337} (\bibinfo{year}{2004}), \eprint{hep-th/0404075}.

\bibitem[{\citenamefont{Silverstein}(2004)}]{Silverstein:2004id}
\bibinfo{author}{\bibfnamefont{E.}~\bibnamefont{Silverstein}},
  \bibinfo{journal}{TASI / PiTP / ISS lectures on moduli and microphysics}
  (\bibinfo{year}{2004}), \eprint{hep-th/0405068}.

\bibitem[{\citenamefont{Grana}(2005)}]{Grana:2005jc}
\bibinfo{author}{\bibfnamefont{M.}~\bibnamefont{Grana}}
  (\bibinfo{year}{2005}), \eprint{hep-th/0509003}.

\bibitem[{\citenamefont{Cremmer et~al.}(1979)}]{Cremmer:1978hn}
\bibinfo{author}{\bibfnamefont{E.}~\bibnamefont{Cremmer}} \bibnamefont{et~al.},
  \bibinfo{journal}{Nucl. Phys.} \textbf{\bibinfo{volume}{B147}},
  \bibinfo{pages}{105} (\bibinfo{year}{1979}).

\bibitem[{\citenamefont{Veneziano and Yankielowicz}(1982)}]{Veneziano:1982ah}
\bibinfo{author}{\bibfnamefont{G.}~\bibnamefont{Veneziano}} \bibnamefont{and}
  \bibinfo{author}{\bibfnamefont{S.}~\bibnamefont{Yankielowicz}},
  \bibinfo{journal}{Phys. Lett.} \textbf{\bibinfo{volume}{B113}},
  \bibinfo{pages}{231} (\bibinfo{year}{1982}).

\bibitem[{\citenamefont{Gibbons}(1984)}]{Gibbons:1984kp}
\bibinfo{author}{\bibfnamefont{G.~W.} \bibnamefont{Gibbons}}
  (\bibinfo{year}{1984}), \bibinfo{note}{three lectures given at GIFT Seminar
  on Theoretical Physics, San Feliu de Guixols, Spain, Jun 4-11, 1984}.

\bibitem[{\citenamefont{de~Wit et~al.}(1987)\citenamefont{de~Wit, Smit, and
  Hari~Dass}}]{deWit:1987xg}
\bibinfo{author}{\bibfnamefont{B.}~\bibnamefont{de~Wit}},
  \bibinfo{author}{\bibfnamefont{D.~J.} \bibnamefont{Smit}}, \bibnamefont{and}
  \bibinfo{author}{\bibfnamefont{N.~D.} \bibnamefont{Hari~Dass}},
  \bibinfo{journal}{Nucl. Phys.} \textbf{\bibinfo{volume}{B283}},
  \bibinfo{pages}{165} (\bibinfo{year}{1987}).

\bibitem[{\citenamefont{Maldacena and Nunez}(2001)}]{Maldacena:2000mw}
\bibinfo{author}{\bibfnamefont{J.~M.} \bibnamefont{Maldacena}}
  \bibnamefont{and} \bibinfo{author}{\bibfnamefont{C.}~\bibnamefont{Nunez}},
  \bibinfo{journal}{Int. J. Mod. Phys.} \textbf{\bibinfo{volume}{A16}},
  \bibinfo{pages}{822} (\bibinfo{year}{2001}),
  \eprint[http://arXiv.org/abs]{hep-th/0007018}.

\bibitem[{\citenamefont{Giddings and Maharana}(2005)}]{Giddings:2005ff}
\bibinfo{author}{\bibfnamefont{S.~B.} \bibnamefont{Giddings}} \bibnamefont{and}
  \bibinfo{author}{\bibfnamefont{A.}~\bibnamefont{Maharana}}
  (\bibinfo{year}{2005}), \eprint{hep-th/0507158}.

\bibitem[{\citenamefont{Brustein and de~Alwis}(2004)}]{Brustein:2004xn}
\bibinfo{author}{\bibfnamefont{R.}~\bibnamefont{Brustein}} \bibnamefont{and}
  \bibinfo{author}{\bibfnamefont{S.~P.} \bibnamefont{de~Alwis}},
  \bibinfo{journal}{Phys. Rev.} \textbf{\bibinfo{volume}{D69}},
  \bibinfo{pages}{126006} (\bibinfo{year}{2004}), \eprint{hep-th/0402088}.

\bibitem[{\citenamefont{Choi et~al.}(2004)\citenamefont{Choi, Falkowski,
  Nilles, Olechowski, and Pokorski}}]{Choi:2004sx}
\bibinfo{author}{\bibfnamefont{K.}~\bibnamefont{Choi}},
  \bibinfo{author}{\bibfnamefont{A.}~\bibnamefont{Falkowski}},
  \bibinfo{author}{\bibfnamefont{H.~P.} \bibnamefont{Nilles}},
  \bibinfo{author}{\bibfnamefont{M.}~\bibnamefont{Olechowski}},
  \bibnamefont{and} \bibinfo{author}{\bibfnamefont{S.}~\bibnamefont{Pokorski}},
  \bibinfo{journal}{JHEP} \textbf{\bibinfo{volume}{11}}, \bibinfo{pages}{076}
  (\bibinfo{year}{2004}), \eprint{hep-th/0411066}.

\bibitem[{\citenamefont{Burgess et~al.}(2003)\citenamefont{Burgess, Kallosh,
  and Quevedo}}]{Burgess:2003ic}
\bibinfo{author}{\bibfnamefont{C.~P.} \bibnamefont{Burgess}},
  \bibinfo{author}{\bibfnamefont{R.}~\bibnamefont{Kallosh}}, \bibnamefont{and}
  \bibinfo{author}{\bibfnamefont{F.}~\bibnamefont{Quevedo}},
  \bibinfo{journal}{JHEP} \textbf{\bibinfo{volume}{10}}, \bibinfo{pages}{056}
  (\bibinfo{year}{2003}), \eprint{hep-th/0309187}.

\end{thebibliography}

\end{document}